\documentclass[letterpaper,twocolumn,10pt]{article}
\pdfoutput=1

\usepackage{usenix-2020-09}

\usepackage{mathtools}
\usepackage{tabularx}
\usepackage{amsmath}
\usepackage{amssymb}
\usepackage{enumitem}
\usepackage{tikz}
\usepackage{ifthen}
\usepackage{caption}
\usepackage{subcaption}
\usepackage{longtable}
\usepackage{booktabs}
\usepackage{listings}
\usepackage[normalem]{ulem}
\usepackage{latexsym}
\usepackage{adjustbox}
\usepackage{fixltx2e}
\usepackage{multicol}
\usepackage{multirow}
\usepackage{framed}
\usepackage{nameref}
\usepackage{amsthm}
\usepackage{url}
\usepackage{float}
\usepackage{setspace}
\usepackage{macrodefs}

\usetikzlibrary{patterns}
\usetikzlibrary{trees}

\title{\codename: A Capability-based Foundation for Trustless Secure Memory
Access (Extended Version)}

\author{
{\rm Jason Zhijingcheng Yu}\\
National University of Singapore
\and
{\rm Conrad Watt}\\
University of Cambridge
\and
{\rm Aditya Badole}\\
National University of Singapore
\and
{\rm Trevor E. Carlson}\\
National University of Singapore
\and
{\rm Prateek Saxena}\\
National University of Singapore
}

\begin{document}

\maketitle

\begin{abstract}

Capability-based memory isolation is a promising new architectural
    primitive. Software can access low-level memory only via
    capability handles rather than raw pointers, which provides a
    natural interface to enforce security restrictions.
Existing architectural capability designs such as CHERI
    provide spatial safety, but fail to extend to other memory models
    that security-sensitive software designs may desire. 
In this paper, we propose \codename{}, a more expressive architectural capability design 
    that supports multiple existing memory
    isolation models {in a \emph{trustless} setup, i.e.,}
    without {relying on trusted software components}. We show how \codename{} is well-suited for environments
    where privilege boundaries are fluid (dynamically extensible),
    memory sharing/delegation are desired both temporally and
    spatially, and where such needs are to be balanced with availability concerns. 
    \codename{} can also be implemented efficiently.
    We present an implementation sketch and through evaluation show that
    its overhead is below $50\%$ in common use cases.
We also prototype a functional emulator for \codename{} and
use it to demonstrate the
    \emph{runnable implementations} of six real-world memory models
    without trusted software components: three types of enclave-based
    TEEs, a thread scheduler, a memory allocator, and Rust-style
    memory safety---all within the interface of \codename{}.

\end{abstract}

\section{Introduction}

Hardware isolation primitives for privilege separation 
play a fundamental role in security designs.
Several security extensions to the memory access interfaces provided
by commodity processors have been proposed and deployed. For example,
trusted execution environments (TEEs)
support ``enclaves'' which are isolated memory regions
accessible only to certain user applications, but not to
privileged software~\cite{sgx, sgxexplained, keystone}. TrustZone~\cite{armtrustzone} partitions all software,
including privileged software, into separate secure and normal worlds. Similarly,
extensions that improve spatial memory safety, e.g., via pointer
integrity (e.g., ARM PAC~\cite{pac,pac-it-up}), bounds checks (e.g., Intel MPX~\cite{intelmpx,
mpk}), and so on, have emerged. 

While each of those extensions is individually promising,
ultimately they are each designed for achieving rather specialized and
rigid forms of memory access restrictions. They cannot be easily
configured for achieving memory protections substantially different
from their original intent efficiently. This makes it difficult for
hardware architecture designers to pick between security extensions to
support natively, which in turn leads to \emph{splintering}: Different
architectures support different security extensions, so software
protections cannot rely on the availability of most of them
ubiquitously. We therefore ask: \emph{Can a hardware-based memory
access model enable several existing memory isolation and
protection models simultaneously?} Such ``one for many'' memory access
models would be a natural solution to splintering.

Two common paradigms for enforcing isolation of memory accesses exist:
access control~\cite{multics,burroughs} and
capabilities~\cite{mondrian,hardbound,mmachine,cheri}. Hardware-based
memory protections based on the classical access control paradigm,
where a security monitor enforces the access policy (read, write,
execute) on every access, are ubiquitous. For example,
privilege rings, enclaves, segmentation, virtualization extensions are
all based on access checks during address translation via memory
management units (MMUs) or memory protection units (MPUs).
However, the access control paradigm requires
tracking a currently executing authority (e.g.,
current privilege ring) and an explicit access
policy for that authority. In contrast, capability-based designs allow
any software to
access memory if and only if it
presents a \emph{capability}, an unforgeable token which grants
its holder rights to access a specific memory region. 
A memory region cannot be accessed without a capability,
which reinforces the
principle of least privilege and reduces ambient
authorities~\cite{confused-deputy, cap-myths}. 
Capabilities do not
require explicit per-context access
control policies---they are implicit in how
accessors transfer capabilities between each other.

CHERI is an example of a capability-based architecture~\cite{cheri}. 
In CHERI,
each capability encodes both
the bounds of the memory region to access and
the types of permitted operations on it (e.g., read, write, execute).
Software can only create
new capabilities from existing ones in a \emph{monotonic} way.
This means newly created capabilities 
cannot grant access rights beyond those of the currently held ones.
CHERI capabilities already provide fine-grained
spatial memory safety: All memory accesses 
are bounded by the capabilities a piece of software
holds. 
This feature is readily useful in software fault isolation or
memory bounds-checking~\cite{stktokens,cheri-provenance,
cheri-cap-model, hardbound}.

However, CHERI-style capabilities are limited in their power to
enforce memory safety in different scenarios.
Firstly, software can use such capabilities in the same way as normal
pointers, creating aliases in different locations. This can
lead to \emph{temporal} safety violations, wherein code forgets to
clean up some capabilities pointing to sensitive data, increasing
the chance of a \emph{capability leak}~\cite{js-cap-leak}.
Second, delegating capabilities across trust boundaries
temporarily is inherently unsafe in CHERI-style
capabilities.
After delegating a capability temporarily to a component,
software has no way to ensure that it no more has access to the memory
region, since it can make copies of the received capability.
In general,
CHERI-style capabilities only allow for
\emph{irrevocable delegation}. 
Third, CHERI-style capabilities do not directly provide
\emph{exclusivity}, i.e., the holder of a capability is not guaranteed
exclusive access to the memory region. However, exclusivity is often needed,
for example, in TEEs~\cite{sgx,sgxexplained,keystone,armtrustzone} and for 
executing critical sections in shared memory
systems~\cite{asyncshock}.

\paragraph{Our approach} We present \codename{}, a new 
capability-based low-level memory access interface intended as an
instruction set architecture (ISA) extension. \codename{} shares the
basic notion of capabilities with CHERI, but adds a novel combination
of improvements which enable it to support many more memory
isolation models at the architectural level---a step towards the goal
of avoiding splintering in systems with finer-grained privilege separation.

Consider an abstract model of a capability machine which runs
$N$ security domains. Each security domain has a register file that
holds
data and capabilities, and can choose to pass them to others through
shared memory regions.
A security domain can also create a new domain by specifying its
initial state and supplying the necessary data and capabilities.
\codename{} provides the following 
{security properties}.
%
%
%
%
%
Firstly, capabilities can be
\emph{linear}\footnote{Linear as in ``linear type systems'' and
``linear logic''.}~\cite{stktokens,cheri,linear-logic,linear-type}.
Beyond granting memory access permissions like ordinary capabilities,
linear capabilities are guaranteed to be \emph{alias-free}, meaning
that no other capability grants a set of memory accesses that overlaps
with that of a linear capability. Therefore, a domain that holds a linear
capability in a register can be sure that it has
exclusive access to the associated memory region. 
Linear capabilities can be derived from one another through spatial
split and merge.
Secondly, \codename{} enables \emph{revocable delegation} of capabilities
across trust boundaries. If $D$ holds a linear
capability and passes it to $E$, $D$ can also choose to revoke this
capability at any time.
$D$ can be assured that $E$ has no access to the memory associated
with the capability immediately after revocation.
This prevents $E$ from keeping its
access permissions to memory indefinitely, or leaking
capabilities by keeping extra copies of them.
Thirdly, \codename{} supports an \emph{extensible hierarchy} of
privileges. Linear capabilities to regions containing other such
capabilities and the above properties hold transitively. Let us say
$D$ holds a capability $c_1$ in a register and the memory region
corresponding to $c_1$ contains a capability $c_2$, which in turn
contains capabilities $c_3$, and so on. $D$ can ensure that it has
exclusive access to all memory regions corresponding to $c_1, c_2,
..., c_n$, can delegate access to any suffix of the chain of such
capabilities and/or immediately revoke access to all delegated
capabilities at once if it so desires.
This considerably simplifies the management of sharing and delegation of
memory, and minimizes the risk of temporal safety bugs or capability
leaks.

\paragraph{Applications}
We present a proof-of-concept prototype of \codename{}
consisting of an ISA emulator and a
compiler for a language with a C-like syntax. 
We demonstrate \emph{with
runnable implementations} how \codename{} can express 
multiple memory isolation or protection
models without relying
on trusted software components.
We implemented three different TEE models: spatially-isolated
enclaves~\cite{keystone, sgx, sgxexplained}, temporally-isolated
enclaves~\cite{elasticlave}, and nested
enclaves~\cite{nested-enclave}.
A spatially-isolated enclave resembles an Intel SGX
enclave~\cite{sgx,sgxexplained}: It has a private memory region
accessible only to itself, and a public memory region also accessible
to the operating system (OS).
The boundaries of those regions are fixed upon enclave creation.
Temporally-isolated TEEs allow dynamic adjustment of access
permissions of memory regions to different enclaves.
This provides a means for secure and efficient memory sharing across
enclaves~\cite{elasticlave,pie,penglai}.
Nested enclaves follow a hierarchical structure, wherein an enclave can
create enclaves inside itself and exchange data with its parent enclave
through a shared memory region~\cite{nested-enclave}. 
{\codename{} supports all those demonstrated TEE models
\emph{trustlessly}, i.e., without involving any trusted software component.}
Most notably, enclaves do not need to trust the memory allocator or
thread scheduler incorporated in our implementations.

\codename{} is also useful in non-enclave applications. Recall
that CHERI capabilities provide spatial safety directly, but not temporal
safety. On \codename{}, we show that one can mimic a Rust-style
ownership and delegation model to achieve both forms of memory safety.
Our implementation enforces such restrictions
through the correct use of
capabilities during runtime, rather than through static type checking,
offering a dynamic alternative for achieving memory safety.

{We formally define the operational semantics of \codename{}.}
{To analyse its security,}
we define an abstract model {that trivially
provides the desired properties (see ``Our approach'' above),
and prove that \codename{} refines it and
also provides the properties therein.}
Our main focus is on the superior \emph{expressiveness} of
\codename{} to support several desirable memory isolation models at
once as compared to CHERI.

\paragraph{Implementation and evaluation}
In order to show that \codename{} can be implemented with acceptable
performance impact, we describe a sketch of a potential
implementation, model its performance with gem5~\cite{gem5}, and
evaluate it on the SPEC CPU 2017 intspeed benchmarks~\cite{spec17}.
The results suggest that overall performance overhead of \codename{}
over a traditional system is within $50\%$.
A full hardware implementation requires
additional design decisions and remains promising future
work beyond the scope of this paper.

\paragraph{Contributions} We present \codename{}, the first architectural
capability design that provides exclusivity, delegation, and
revocation simultaneously for hardware-isolated memory. 
\codename{} enables richer memory models demanded by security
applications with a single set of interfaces than prior capability-based
systems.

\section{Overview}

Existing memory isolation models require specialized architectural
support. While this enables efficient
implementations ultimately, the plurality of such models has led to
splintering. \codename{} is an effort towards finding abstractions
expressive enough to be configured to support multiple useful
existing isolation models without increasing the software trusted
computing base (TCB).

\subsection{Architectural Capabilities}
A capability is a token that grants its holder
memory access permissions.
It typically contains the bounds of the
accessible memory locations as well as the allowed access types (i.e., read, write,
execute).
Software presents a capability every time it needs to make a memory
access.
The hardware then performs checks on the memory access to make sure 
that it falls in the allowed bounds and is of an allowed type of the
capability.
Whenever the memory access is found to violate the restrictions, the
hardware refuses to fulfil the memory access.
Capabilities are unforgeable---software can only derive
new capabilities from existing ones through a well-defined set of
operations.
For example, software cannot directly cast an integer into a
capability.
Implementations of capability-based architectures like CHERI enforce
this through memory tagging~\cite{hardbound,mmachine,cheri} by marking
memory locations and registers containing capabilities with tags
that are hidden from software.
The operations that create capabilities out of existing ones
are \emph{monotonic}, i.e., new capabilities cannot
allow accesses disallowed by the original ones.
This prevents privilege
escalation through direct operations on a capability.

Compared to identity-based access control mechanisms, capabilities
have the advantage of not relying on complex central
policies, and can yield greater expressiveness.
As an example,
CHERI~\cite{cheri,cheriabi} has been shown to enable
fine-grained software compartmentalization and spatial memory safety
in C/C++.
However, it does not provide architectural support for temporal
memory safety.
To enable temporal memory safety for C/C++, for example, traditional
software-based techniques such as reference counting and garbage
collection must be used in conjunction~\cite{cheri-memory-safe-c}.
For several other application scenarios, extensions to CHERI that
require specialized hardware changes exist.
Such examples include StkTokens~\cite{stktokens}, which enables a
calling convention that guarantees well-bracketed control flow in
software fault isolation, and
CHERIvoke~\cite{cherivoke} and Cornucopia~\cite{cornucopia}, which
mitigate use-after-reallocation of heap memory for C code.

\subsection{Motivating Examples}
\label{subsec:examples}
Many memory isolation models are useful in the real world
but are not supportable with CHERI, motivating our work.

\paragraph{Trustless memory allocation}
One important task of the OS and the VMM is the allocation of physical memory.
Traditionally, an application has to trust privileged code
when using the allocated memory.
Achieving trustless memory allocation requires considering two
aspects.
Firstly, the application that receives an allocated memory region from the memory
allocator should not trust that it will not
access the memory region or delegate it to another
application in the future.
Secondly, the memory allocator should not overly trust 
applications, which may refuse to relinquish access to memory regions.

CHERI is unable to achieve trustless memory allocation.
When the application receives a capability from the memory allocator, it cannot ensure
that no other software component, including the memory allocator,
also has
access to the allocated memory region.
Likewise, the memory allocator cannot be sure that the
application has relinquished the capability when it wishes to
reclaim it. 
The memory allocator needs to trust that the application has not kept or leaked
copies of the reclaimed capability.
As a result,
both the allocator
and the application would need to
trust each other.

\paragraph{Trustless preemptive scheduling}
Modern systems widely rely on preemptive scheduling to multiplex
multiple domains (e.g., processes) on limited CPU resources.
Preemptive scheduling relies on preempting (i.e., interrupting) the
execution of a domain through timer interrupts.
A scheduler then handles each interrupt and decides which domain to
execute next.
The scheduler is normally part of an OS and has
the privilege to arbitrarily access domain execution states.
On the contrary, trustless preemptive scheduling enforces the \emph{principle of least
privilege} and provides
applications with the assurance that
the scheduler is not capable of doing anything more than deciding
when to execute each domain.
This effectively removes the scheduler from the TCB of an application.

On CHERI, an exception or interrupt on a thread diverts the control
flow to an exception handler, which can then perform scheduling.
However, the execution context of the interrupted thread, including
all the capabilities stored in registers, is directly accessible to
the exception handler.
This gives the scheduler the
ability to modify the content of the execution context
of a domain (e.g., register values), for example,
to hijack the domain control flow. 
The scheduler can also duplicate the execution context and force
application code
that is not designed to be thread-safe to interleave on multiple
threads on shared memory regions through capabilities duplicated as
part of the execution context.
Therefore, CHERI requires the application
to fully trust the scheduler.

\paragraph{Trusted execution environments}
Traditionally, software such as OS kernels is assumed to be trustworthy
and runs with high privileges.
However, the growing complexity of privileged software and the
increasing demand for secure remote execution
have rendered this assumption
increasingly unjustifiable.
Trusted execution environments (TEEs) provide a promising solution: 
They support running
security-sensitive software
without requiring it to trust any other software on the system,
including 
privileged software such as the OS.
Most TEEs follow a \emph{spatial isolation model}, where each
secure application receives
a private memory region called
an \emph{enclave} at its launch time.
An enclave 
stores both the code 
and the private data of the application,
and is accessible only to it.
The remaining part of the memory, called the \emph{public memory} and
accessible to both the application and the OS, enables
data exchange between them (e.g., to support
system calls).
Variations of the enclaved TEE model exist.
In the \emph{nested
enclave} model~\cite{nested-enclave}, for example,
an application in a nested enclave has access to its parent
enclave in the same way as how an application in a top-level enclave
has access to a public memory shared with the OS. 
Another variant, Elasticlave~\cite{elasticlave}, supports temporal
isolation, where each application can set time-varying access policies for
its memory regions for sharing them in a controlled way.
Both the nested enclave and the Elasticlave models enable greater
flexibility and a wider range of application scenarios than
spatial isolation.

CHERI does not support any of the above-mentioned TEE models, as it
does not guarantee exclusive access for applications.
Any memory region 
an application can access through a
capability, even one intended as an enclave private memory region,
can potentially be accessible to other software
components as well,
by passing them copies of the same capability.

\paragraph{Rust-like memory restrictions}
Many applications involve sharing memory across software components.
An example is a Linux process sharing a buffer with the kernel
in order to read data from a file.
In such cases, it is important to maintain spatial and temporal safety
of memory access across domains, and violations can lead to serious
consequences.
Memory-safe abstractions are one answer to this problem.
Rust~\cite{rust}, a programming language that provides a
memory model with spatial and temporal safety
is an example of such models.
However, abstractions which
rely on static enforcement by compilers require that software
components be written in specific languages and require additional trust
assumptions, i.e., that components trust each other to have used
a correct compiler implementation without taking unsafe shortcuts.

Rust enforces memory safety with the notions of ownership
and lifetime.
Rust programs access data objects through their references.
Though an object can have more than one reference, exactly one of them
is its \emph{owner}, while the others are all \emph{borrowed
references}.
Hence, 
the owner reference of an object can only be \emph{moved}, but not
\emph{duplicated}.
The owner's lifetime is tied to that of the object, and
a borrowed reference cannot outlive the owner.
This makes sure that no reference to an object can exist after the
object is destroyed (i.e., when the
owner's lifetime ends).
It also implicitly guarantees that 
access to an object will be exclusive to the owner
again after the borrowed references are destroyed.

The CHERI capability interface, however, is not expressive enough to
directly enable Rust-style memory restrictions at the architectural level.
Rust involves different types of references and imposes
different restrictions on them. 
CHERI, on the other hand, provides only one type of capabilities.
It also does not provide revocable delegation, as exemplified by
borrowed references in Rust.
Once a domain delegates certain memory access to another domain,
there is no guarantee that it can get back exclusive memory
access at a future point.

\subsection{\codename{} in a Nutshell}
\label{subsec:goals}

We design a new memory access interface
called \codename{} that can express the memory restrictions needed by
the memory models discussed in Section~\ref{subsec:examples}.
\codename{} uses capabilities for
memory access control in the physical address space.
As an architecture-level interface, it is intended to be
implemented in the processor, with capabilities replacing raw memory
addresses.
The processor enforces
memory protection guarantees at runtime without assuming trusted
software components.
\codename{} does \emph{not} rely on assumptions regarding the MMU
on a system by directly working with physical
addresses instead of virtual ones.
In the future,
this could eventually enable greater flexibility and
compatibility with isolation models not built
on MMUs or MPUs (e.g., ARM MPU~\cite{arm-mpu}, RISC-V PMP~\cite{riscvpriv}).
On top of 
the original capabilities from CHERI,
\codename{} adds new capability types
with the following properties:

        \emph{(P1) Linearity.} Domains can have exclusive
        access to memory regions. When a domain $D$ holds a linear
        capability to a memory region, no other domain can access the region.

        \emph{(P2) Delegation and revocation.} When a domain $D$ holds a
    linear capability $L$, $D$ may choose to transfer $L$
        to another domain $E$.
Moreover, $D$ may later choose to reclaim exclusive ownership of $L$,
        even if $E$ has in
        turn transferred the capability to another domain.
To protect the potentially secret data that $E$ may have placed in
        locations accessed through $L$, when $D$ regains ownership of
        $L$, 
       
        the memory region corresponding to $L$ will become
        unreadable to $D$ until $D$ overwrites it.

        \emph{(P3) Dynamically extensible hierarchy.} 
    A domain $D$ can create another domain $E$ that is
        \emph{subordinate} to it, in the sense that
        $D$ can choose to revoke \emph{any} capability
        that $E$ holds at \emph{any} time, and once this is done, $E$
        cannot get back the revoked capability without $D$'s
        cooperation.
        Such a hierarchy is dynamically defined by the
        runtime behaviour of each security domain and can be
        indefinitely extended on demand.

        \emph{(P4) Safe domain switching.}
    If at a certain moment the physical thread executing a domain $D$
        switches to a different domain 
        (e.g., due to an exception/interrupt, or when
        calling into another domain), and $D$'s
        context (register file content) is $\mathcal{C}$, then the
        next time $D$ is executed, its context is still $\mathcal{C}$.

\begin{table}[t]
    \caption{Properties required or present in each model.}
    \label{table:examples-properties}
    \begin{center}
        {\small
    \begin{tabular}{lcccc}
        \hline
        \textbf{Model} & \textbf{P1} &
        \textbf{P2} & \textbf{P3} & \textbf{P4} \\
        \hline
        Rust-like abstraction~\cite{rust} & \tableyay{} & \tableyay{} &
        \tablenay{} & \tablenay{} \\
        Spatially-isolated enclaves~\cite{sgx,sgxexplained} & \tableyay{} & \tableyay{} &
        \tablenay{} & \tableyay{} \\
        Temporally-isolated enclaves~\cite{elasticlave} & \tableyay{} & \tableyay{} &
        \tablenay{} & \tableyay{} \\
        Nested enclaves~\cite{nested-enclave} & \tableyay{} & \tableyay{} & \tableyay{} & \tableyay{} \\
        Trustless memory allocation & \tableyay{} & \tableyay{} &
        \tablenay{} & \tablenay{} \\
        Trustless thread scheduling & \tableyay{} & \tablenay{} &
        \tablenay{} & \tableyay{} \\
        \hline
        CHERI~\cite{cheri} & \tablenay{} & \tablenay{} & \tablenay{} & \tablenay{} \\
        \codename{} (this work) & \tableyay{} & \tableyay{} & \tableyay{} &
        \tableyay{} \\
        \hline
    \end{tabular}}
    \end{center}
\end{table}

Table~\ref{table:examples-properties}
{lists}
the properties the example models require.

\paragraph{\codename{} abstract model \& security}
To
{capture} 
our security properties more precisely,
we define an abstract model called \abscodename{}.
{Its state} is defined in terms of an abstract memory
store, where memory cells may be marked \textsf{uninit} to capture
that reading them would be a security violation, together with a
\textit{user domain} executing in a two-part environment composed of
the \textit{superordinate} 
and the \textit{subordinate}
environment domains ($\textit{tstate}_\textsf{sup}$ and $\textit{tstate}_\textsf{sub}$).
The superordinate environment represents other domains which may
revoke the user domain's capabilities arbitrarily.
The subordinate environment represents domains which are guaranteed to
never revoke the user domain's linear capabilities.
The user domain and the two environments each track the memory accessible to them through the capabilities
they currently own, and can perform actions
($\textit{act}_\textit{abs}$) to manipulate these capabilities or
update the memory.
\abscodename{} directly enforces desirable properties of \codename{}, and we characterize \codename{}'s security as a standard refinement theorem:

\begin{mainthm*}
    \label{thm:main}
    \codename{} refines \abscodename{}.
\end{mainthm*}

We discuss
\abscodename{} and the proof of the main theorem in
Section~\ref{sec:security}.

\subsection{Threat Model and Scope}
We assume that the security domains do \emph{not} trust
one another.
We focus on the security of one of the security domains, and
assume that the attacker can control any other domain on the
system, including those in charge of managing system
resources (e.g., thread scheduler, heap memory allocator, and so on). 
The domain of interest, on the other hand, is assumed to be
benign and bug-free.
We are also interested in denial-of-service (DoS) attacks from an application
which attempts to hold memory resources indefinitely and thereby
reject them to OS components in charge of memory management.
DoS attacks from the OS against applications are out of scope.

\begin{figure}[t]
\centering
\includegraphics[width=\linewidth]{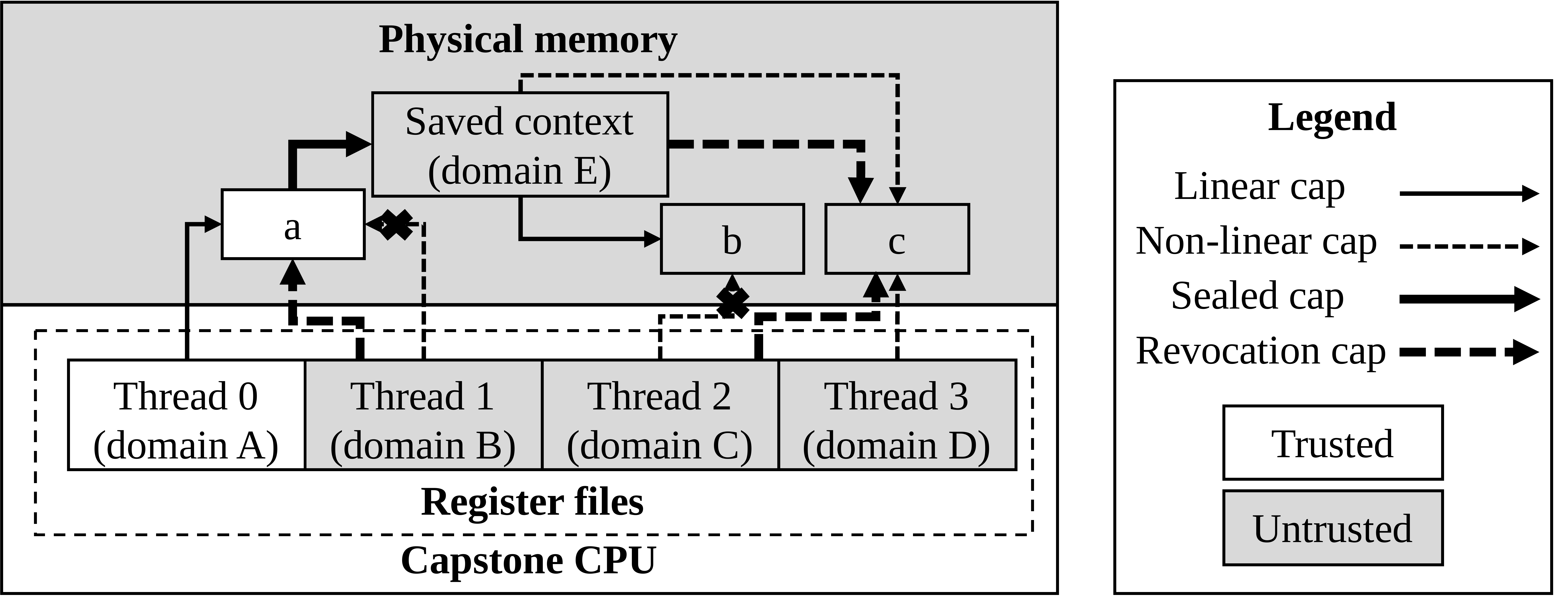}
    \caption{Overview of \codename{}. An arrow from $X$ to
    $Y$
    represents a capability located inside $X$ that grants access
    to $Y$. Crossed-out arrows are capabilities
    not allowed to exist.}
\label{fig:linear-cap-example}
\end{figure}

\section{Design Overview}
\label{sec:design}

A \codename{} machine runs multiple \emph{security domains} multiplexed on
a set of physical threads which have separate
register files but share the physical memory (and the physical address
space), as shown in Figure~\ref{fig:linear-cap-example}.
At any given point in time, each thread is running \emph{exactly} one
security domain, and each security domain is running on \emph{at most} one
physical thread. 
Certain events (e.g., exceptions) can trigger a thread to switch
between security domains.
When a security domain is running on a physical thread, we refer to
the register file content of the thread as the
\emph{context} of the security domain.
For a security domain that is not currently running on any physical
thread, we define its context as the context it will have when it next
starts running.
Such a context is physically stored inside a memory region, from which
the content is loaded into the register file of a thread when
it loads the security domain.

\begin{figure}[t]
    \centering
    \resizebox{.9\linewidth}{!}{
    \begin{tikzpicture}[yscale=0.55,xscale=0.95] 
        \node[draw, rounded corners=2pt, fill=black, text=white](linear) at (0, 0) {Linear};
        \node[draw, rounded corners=2pt, fill=diagramgray](non-linear) at (3.3, 0) {Non-linear};
        \node[draw, rounded corners=2pt](revocation) at (0, 3.3) {Revocation};
        \node[draw, rounded corners=2pt, fill=black, text=white](sealed-ret) at (-3.3, 3.3) {Sealed-return};
        \node[draw, rounded corners=2pt, fill=black, text=white](uninitialized) at (3.3, 3.3) {Uninitialized};
        \node[draw, rounded corners=2pt, fill=black, text=white](sealed) at (-3.3, 0) {Sealed};
        \draw[-stealth] (linear) -- (non-linear) node [midway, above]
        {delinearize};
        \draw[-stealth] (linear) to[out=110, in=-110] node [midway, sloped,
        below] {mint
        rev} (revocation);
        \draw[-stealth] (revocation) to[out=-70, in=70] node [midway,
        sloped, below]
        {revoke} (linear);
        \draw[-stealth] (revocation) -- (uninitialized) node [midway, above]
        {revoke};
        \draw[-stealth] (linear) -- (sealed) node [midway, above] {seal};
        \draw[-stealth] (uninitialized) -- (linear) node [midway,
        right]
        {initialize};
        \draw[-stealth] (sealed) to[out=110, in=-110] node [midway, sloped,
        below] {call} (sealed-ret);
        \draw[-stealth] (sealed-ret) to[out=-70, in=70] node [midway,
        sloped, below]
        {retseal} (sealed);
    \end{tikzpicture}}
    \caption{Overview of different types of capabilities in
    \codename{} and the operations that change the type of a capability.
    Capability types with black backgorunds are alias-free,
    non-linear capabilities can overlap among themselves, whereas
    revocation capabilities can overlap with any other capabilities.
    }
    \label{fig:cap-type-change}
\end{figure}
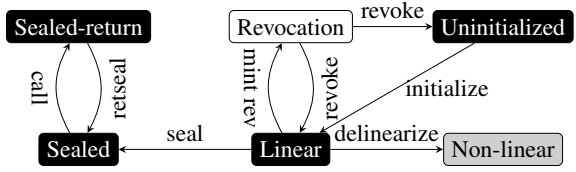

Similar to CHERI,
\codename{} is an instruction set architecture (ISA) 
based on capabilities.
For \emph{any} memory access by \emph{any} security domain,
\codename{} mandates that a capability granting this memory access
must be provided.
Each domain context can
hold capabilities that grant the domain access to memory regions,
which can in turn hold more capabilities and hence
grant access to yet more memory regions.
To overcome the limitations of existing
capability-based architectures (Section~\ref{subsec:examples}),
\codename{} incorporates extra capability types and
capabilities-related operations
beyond those already present in CHERI.
Some capability-related operations involve
changing capability types.
Figure~\ref{fig:cap-type-change} overviews the
capability types and the operations. 
We describe them in more details below.

\paragraph{Linear capabilities}
Central to \codename{} 
is the additional capability type called \emph{linear
capabilities}~\cite{stktokens,cheri,linear-logic,linear-type}.
A linear capability grants access to memory locations in the same way
as an ordinary capability,
but instead of \emph{only} guaranteeing that certain
memory accesses are \emph{allowed}, linear capabilities also assure
that certain accesses are \emph{disallowed}.
This is because linear capabilities are
\emph{alias-free}.
Holding a linear capability not only gives a domain 
certain access permissions to the memory region, but is also
sufficient to
guarantee that access to the region is exclusive to the domain
alone.
This does not assume any trust in software.

To maintain the alias-free property of linear capabilities,
any operation in \codename{}
that would otherwise lead to overlap between input and output
capabilities will consume (i.e., invalidate)
the input capabilities.
For example,
software can only \emph{move}, but not \emph{copy} linear
capabilities.\footnote{Moving a linear capability from location
$A$ to $B$ destroys the copy in $A$.}

\begin{figure}[t]
    \centering
    \includegraphics[width=\linewidth]{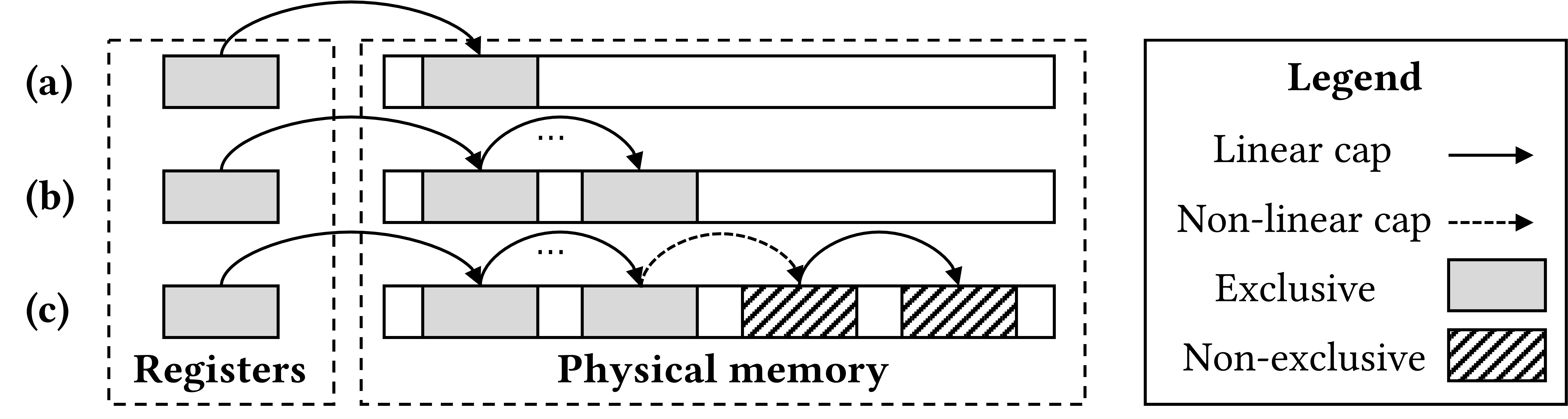}
    \caption{Exclusive access guarantees in different scenarios: (a)
    Through a linear capability; (b) Through a chain of linear
    capabilities; (c) No exclusive access guarantee through a chain
    with non-linear capabilities.}
    \label{fig:cap-chain}
\end{figure}

A linear capability does not need to be in a domain
context to guarantee exclusive access.
For example, when a domain holds in its context a linear capability $c_u$
for the memory region $R_u$, and inside $R_u$ resides
another linear capability $c_v$ for another memory region $R_v$, then
besides $R_u$, the
domain also has exclusive access to $R_v$.
In general, exclusive access through linear capabilities can be
chained indefinitely.
As shown in Figure~\ref{fig:cap-chain}, 
when a domain can reach a memory region through a linear capability
kept directly in its context (register file)
(Figure~\hbox{\ref{fig:cap-chain}(a)}), or through a chain of linear
capabilities (Figure~\hbox{\ref{fig:cap-chain}(b)}), 
its access to the memory region is guaranteed to be exclusive.
On the other hand,
exclusive access is not guaranteed
if the domain has to involve a non-linear
capability to reach the memory region 
(Figure~\hbox{\ref{fig:cap-chain}(c)}).

\begin{figure}[t]
    \centering
    \includegraphics[width=\linewidth]{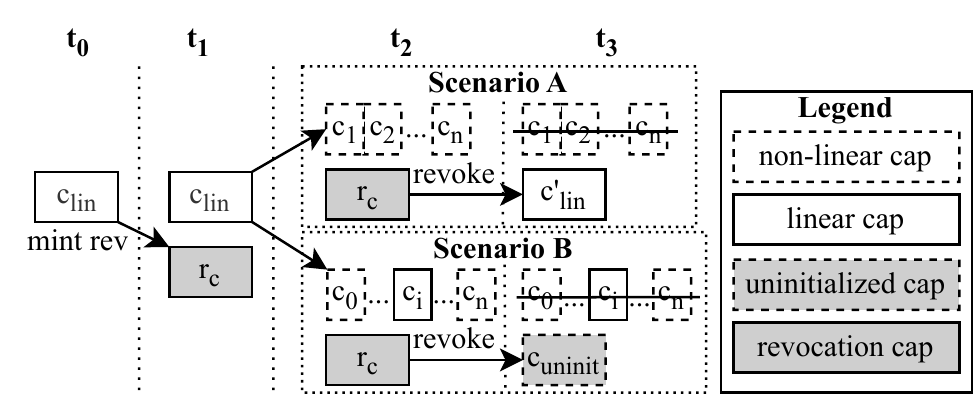}
    \caption{{Overview of the operations on revocation capabilities.
    Strikethrough capabilities are invalid.
    When $c_{\mathrm{lin}}$ has derived non-linear capabilities only at $t_3$, revocation
    converts $r_c$ into a linear capability $c^\prime_{\mathrm{lin}}$ (scenario A).
    Otherwise, $r_c$ is converted into an uninitialized capability $c_{\mathrm{uninit}}$ (scenario B).
    }}
    \label{fig:revocation}
\end{figure}

\paragraph{Revocation}
\codename{} includes the notion of \emph{revocation capabilities}.
A revocation capability does not grant memory access permissions,
but serves as a token for revoking all capabilities that overlap with
it.
{As demonstrated in Figure~\ref{fig:revocation},}
a domain can create a revocation capability only for a
linear capability it currently holds ({$t_0$ to $t_1$ in Figure~\ref{fig:revocation}, 
and} ``mint rev'' in
Figure~\ref{fig:cap-type-change}).
Creating a revocation capability
does \emph{not} consume the given linear
capability. This does not violate its alias-free property, as the
revocation capability conveys no memory access permissions.
As such, the revocation capability serves as a basis for
revocable delegation.
Before a domain $D$ passes a linear capability to another domain
$E$, it
creates a revocation capability for the linear capability,
which it later
uses to revoke the delegated capability,
regardless of what $E$ has done.
In order for $D$ to reclaim exclusive access
to the memory region, \codename{} converts the revocation capability
into a corresponding capability that grants access permissions in the revocation
operation ({$t_2$ and $t_3$ in scenario A in Figure~\ref{fig:revocation},
and} ``revoke'' in Figure~\ref{fig:cap-type-change}).
Since linear or non-linear capabilities that overlap with the memory
region are all revoked, the new capability does not violate the alias-free
property.
On some occasions, \codename{} converts the revocation capability into
an \emph{uninitialized capability} instead of a linear capability
to prevent secret leakage.
For example, 
when $D$ reclaims
exclusive access, the memory region possibly holds $E$'s secret
data.
\codename{} identifies such situations by checking whether a linear
capability has been revoked during the revocation process, 
which indicates a domain is still
interested in maintaining its exclusive access {($t_2$ and $t_3$ in scenario B in
Figure~\ref{fig:revocation})}.
An uninitialized capability represents a memory region
whose content should be unavailable until
written (hence effectively uninitialized).
Correspondingly, an uninitialized capability only grants write access,
but can be converted to a linear
capability when all locations inside the
region have been written at least once with it (``initialize'' in
Figure~\ref{fig:cap-type-change}).
\codename{} thus prevents the domain that
reclaims access to the memory region from reading its original
content.

\begin{figure}[t]
    \centering
    \includegraphics[width=.8\linewidth]{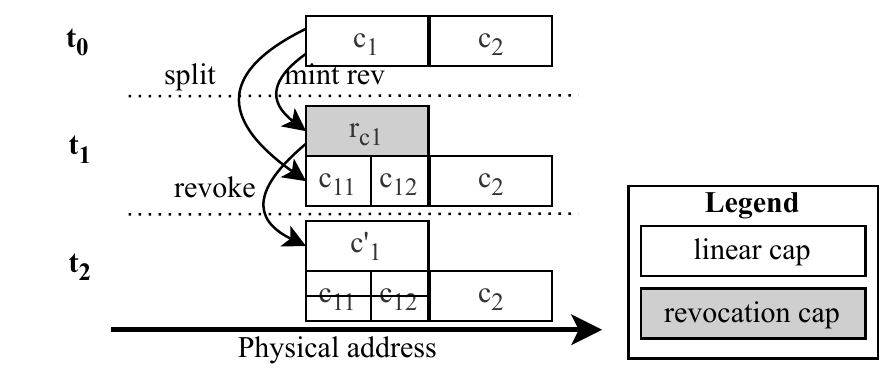}
    \caption{{Split/merge.
    Strikethrough capabilities are invalid.}}
    \label{fig:split-merge}
\end{figure}

\paragraph{Splitting and merging}
\codename{} allows {spatially}
splitting a linear capability into two
non-overlapping linear capabilities.
Since this operation is entirely monotonic,
a reverse process is needed to merge linear capabilities
and stop linear capabilities from becoming
increasingly fragmented.
However, it is infeasible to simply allow any two linear
capabilities for adjacent regions and identical permissions to be
merged.
Consider the scenario where a linear capability $c_1$ has a
corresponding revocation capability $r_{c_1}$, and is later split into
two linear capabilities $c_{11}$ and $c_{12}$.
Another linear capability $c_2$ neighbours $c_{12}$ in terms of their
memory regions, and has identical permissions as $c_{12}$.
If under this condition we merge $c_2$ and $c_{12}$ into a new
linear capability $c_3$,
problems will arise when $r_{c_1}$ is used to perform
revocation: On the one hand, $c_3$ overlaps with $r_{c_1}$, so it
should be revoked; on the other, $c_3$ is not entirely covered by the
memory region associated with $r_{c_1}$, and if it is revoked, the
$c_2$ part of $c_3$ will be lost.
This introduces significant complexity in capability management and
poses challenges to both implementations and applications.
As a result, \codename{} avoids such
arbitrary merging, and instead relies on the semantics of revocation for
the reverse operation of splitting.
Before splitting a linear capability, a domain creates a
revocation capability, and later uses it to revoke the capabilities
that result from this split as well as reclaim the original
linear capability {(Figure~\ref{fig:split-merge})}.
In the example, $c_{12}$ can only be merged with $c_{11}$ to
reverse the split of $c_1$, by performing revocation using $r_{c_1}$.
This enables a limited form of merging that reverses past
splits,
which we consider
as a reasonable compromise between complexity and utility.
{Revocation capabilities are used to reverse other types of operations
as well: for example,
tightening capability permissions and delinearizing linear capabilities (i.e., converting them into
non-linear capabilities, shown as ``delinearize'' in
Figure~\ref{fig:cap-type-change}).}

\begin{figure}[t]
    \centering
    \includegraphics[width=\linewidth]{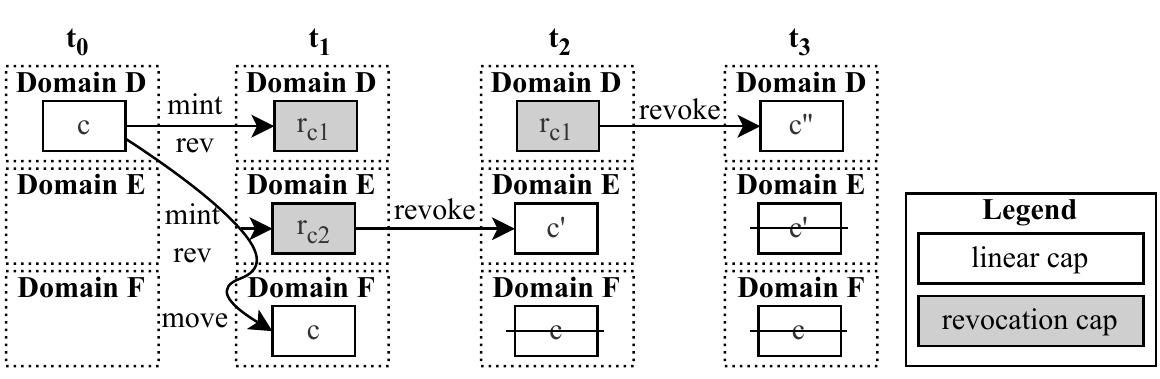}
    \caption{{Extensible capability delegation hierarchy through
    revocation capabilities. Strikethrough capabilities are invalid.}}
    \label{fig:hierarchy}
\end{figure}

\paragraph{Implicit extensible hierarchy}
\codename{} provides an implicit and extensible \emph{hierarchy} through
revocation capabilities.
Revocation capabilities can overlap with one another.
For example, after a domain $D$ creates a revocation capability
$r_{c1}$ for the linear
capability $c$, it passes $c$ to another domain $E$, which in turn
creates another revocation capability $r_{c2}$ for $c$.
$E$ may pass
$c$ further to a third domain $F$ while retaining $r_{c2}$ so it can later
revoke $F$'s access permissions {($t_0$ to $t_1$ in Figure~\ref{fig:hierarchy})}.
In such a case, $r_{c1}$ and $r_{c2}$ should not be considered as
identical.
If $r_{c1}$ is used to perform revocation, $r_{c2}$ should be
revoked, or $E$ will be able to regain $c$ through $r_{c2}$
afterwards, whereas if $r_{c2}$ is used for revocation first {($t_2$ in
Figure~\ref{fig:hierarchy})},
$r_{c1}$ should remain valid, so that $D$ can still revoke $E$'s
access regardless of this event {($t_3$ in Figure~\ref{fig:hierarchy})}.
\codename{} deals with such situations by assigning 
each revocation capability with
a different strength based on seniority: The
earlier a revocation capability was created, the stronger it is.
A revocation operation with a revocation capability $r$ invalidates 
all other \emph{weaker and overlapping} revocation capabilities.
{Note that such other revocation capabilities
can only point to spatial sub-regions (including the identical region) of the memory region
that $r$ points to, because
creating a revocation capability requires a valid
linear capability.}
Such a hierarchy of revocation strengths
is
implicit in how security domains delegate linear
capabilities and is indefinitely extensible.
A linear capability can be passed indefinitely many times across a
sequence of domains.
Each domain can always ``roll back'' passing a linear
capability to the next domain, regardless of the behaviours of the domains
further down the sequence.

Capability-based designs have the advantage that they can work without access control policies written to be enforced by security monitors. It frees us from defining access control policies upfront. In contrast, when capabilities are passed from program context to context, say from one process to another, they implicitly carry with it the semantics that the sender context wishes to allow the recipient access to the object. This implicit capability-passing is a form of delegation without explicit intervention or access control decisions being made. This also means that if we design capability-based models, we do not need to define privilege levels explicitly.

\paragraph{Safe domain switching}
\codename{} supports safe domain switching with the help of 
\emph{sealed capabilities}, which are present also in
CHERI~\cite{cheri}.
Similar to revocation capabilities,
sealed capabilities do not grant direct memory access.
Instead, a sealed capability
represents the context of a security domain that is not
currently running.
The memory region associated with a sealed capability stores the
content of a domain context.
A linear capability can be converted into a sealed
capability (``seal'' in Figure~\ref{fig:cap-type-change}), which
in effect creates a security domain with a specified context.
A domain can use the ``call'' operation on a sealed
capability to switch the current physical thread to the corresponding
domain.
A similar operation is ``return'', which also switches to
a specified domain context stored inside a memory region, but is
intended as the reverse of ``call''.
\codename{} accounts for the semantic difference between ``call'' and
``return'' with a separate sealed-return capability type,
which is generated in the ``call'' operation as shown in
Figure~\ref{fig:cap-type-change}.
Unlike CHERI, \codename{}
extends sealed capabilities
to exception handling as well, guaranteeing
that the exception handler cannot arbitrarily access the execution context
of an interrupted domain.
Furthermore, in \codename{}, sealed capabilities are linear (i.e.,
guaranteedly alias-free), which
ties access to
the stored resources behind a sealed
capability to the domain.
This also guarantees that only one instance of a 
domain exists at any time, effectively preventing potentially
unsafe re-entries into the same domain.

\paragraph{Alias-free capability types}
Linear capabilities are not the only capabilities 
with alias-free guarantees.
As Figure~\ref{fig:cap-type-change} shows, sealed,
sealed-return, and uninitialized capabilities are also
alias-free,
allowed to overlap only with revocation capabilities, whereas
revocation capabilities 
can
overlap with any capability of any type, and
non-linear capabilities 
can
overlap with other non-linear capabilities and revocation capabilities.

{
\paragraph{Software stack}
Upon a system reset, the register file is
initialized to contain
capabilities that cover the full physical memory.
The first piece of code to execute can then bootstrap other domains
and delegate to them parts of the physical memory as linear capabilities.
Multiple paths henceforth are worth exploring,
from adapting a monolithic kernel
such as Linux to creating a new microkernel-based software stack.
Detailed software stack design is future work.
}

\paragraph{Supporting memory protection models}
The design delineated above enables \codename{} to provide
the desired properties discussed in Section~\ref{subsec:goals}, which
as summarized in Table~\ref{table:examples-properties},
are required to support the example memory
protection models but are missing in CHERI.
We describe more details on how to implement those 
models on \codename{} in Section~\ref{sec:app}.

\section{\codename{} Formal Model}
\label{sec:model}

\subsection{Overview}

Figure~\ref{fig:syntax} defines the entities \codename{} involves.

\begin{figure}[t]
    {\small
\begin{tabular}{rcl}
$n, N, d, b, e, a$ & $\in$ & $\fullN$ \\

$\sReg \ni r$ &$\coloneqq$& $\rpc \mid
    \rret \mid \repc \mid \rr_1 \mid \rr_2 \mid \cdots \mid \rr_M$\\

$\sCapType \ni t$ & $\coloneqq$ & $\tLin \mid \tNon \mid \tRev
        \mid \tSealed(d) \mid$\\
&& $\tSealedRet(d, r) \mid \tUninit$\\


$\sPerms \ni p$ & $\coloneqq$ & $\pNA \mid \pR \mid \pRW \mid \pRX \mid \pRWX$\\

$\sRNodeType \ni nt$ & $\coloneqq$ & $\tRLin \mid \tRNon$\\

$\sRevParent \ni pr$ & $\coloneqq$ & $n \mid \nroot \mid \nnull$\\

$\sRevTree \ni rt$ & $\coloneqq$ & $\fullN \mapsto \sRevParent \times
    \sRNodeType$\\

$\sCap \ni c$ & $\coloneqq$ & $\capfd$\\

$\sWord \ni w$ & $\coloneqq$ & $c \mid n \mid i$\\

$\sRegFile \ni regs$ & $\coloneqq$ & $\sReg \mapsto \sWord$\\

$\sMemory \ni mem$ & $\coloneqq$ & $\fullN \mapsto \sWord$\\

$\sThreadState \ni \theta$ & $\coloneqq$ & $regs \mid \aerror{}$\\

$\sThreads \ni \Theta$ & $\coloneqq$ & $\fullN \mapsto \sThreadState
        \times \fullN$\\
        
$\sState \ni \mstate$ & $\coloneqq$ & $\mstatefd$ 
\end{tabular}

$\sInsn \ni i \coloneqq
    \imov{}\ r\ r \mid
    \ild\ r\ r \mid
    \isd\ r\ r \mid
    \itighten\ r\ r \mid
    \ishrink\ r\ r\ r \mid
    \isplit\ r\ r \mid
    \idelin\ r \mid
    \iscc\ r\ r \mid 
    \ilcc\ r\ r \mid
    \imrev\ r\ r \mid
    \idrop\ r \mid
    \iseal\ r \mid
    \icall\ r\ r \mid
    \ireturn\ r\ r \mid
    \iretseal\ r\ r \mid
    \irevoke\ r \mid
    \iinit\ r \mid
    \iexcept\ n \mid 
    \ijmp\ r \mid 
    \ijnz\ r\ r \mid
    \ili\ r\ n \mid
    \iadd\ r\ r \mid 
    \ilt\ r\ r\ r \mid
    \iinvalid 
    $}

\caption{Syntax of the \codename{} model.}
\label{fig:syntax}
\end{figure}

\paragraph{\codename{} machine}
The execution of a \codename{} machine consists of a sequence of
steps.
At each step, the \codename{} machine is in a state $\mstate$
consisting of the states of physical threads $\Theta$,
physical memory state $mem$, and additional capability-related data
structures.
Each physical thread has a distinct register file which includes
the program counter $\rpc$, special registers 
$\rret$ and
$\repc$, and $M$ general-purpose
registers $\rr_1, \rr_2, \cdots, \rr_M$.
By indexing $\Theta$, we can obtain the state $\theta$ of a
specific physical thread, including its register file contents.
Each register or memory location contains either a raw scalar value or
a capability, referred to as a \emph{word} collectively.

At each step, the machine picks \emph{any one} of the threads and
executes an instruction on it.
The instruction can be either the one stored in the physical memory
under the cursor of the capability held by $\rpc$, or $\iexcept$,
a special instruction that helps
model an exception or interrupt.
Executing an instruction changes the machine state.
We represent the machine state immediately 
after executing instruction $i$ on thread $k$ at the machine state
$\mstate$ as $\fExecute(\mstate, k, i)$.

\paragraph{Capabilities}
Each capability is a tuple $c = (t, b, e, a, p, n)$, where $b$
and $e$ are the base and end addresses of the memory region
respectively, $p$ is the access permissions granted by the capability, and
$t$ identifies the capability type.
\codename{} includes new capability types \emph{in
addition} to the normal non-linear capability (denoted as $\tNon$).
We follow CHERI~\cite{cheri} to include in each capability
the address for the next
memory access, called its \emph{cursor} and denoted as $a$.
Information useful for capability revocation is recorded in
$n$.

\paragraph{Operational semantics}
\label{subsec:semantics}
{
We discuss the semantics of each instruction in turn below.
}
The instructions $\ijmp, \ijnz, \ili, \iadd$, and $\ilt$ are omitted,
as they are
almost identical to their counterparts in common existing
architectures.
{
Interested readers may find the full definition of the state transitions
in Appendix~\ref{appx:complete-model}.
}

\subsection{Moving Capabilities}
As in traditional architectures (e.g.,
RISC-V~\cite{riscvpriv}),
$\ild$ and $\isd$ perform memory load and store
operations, but require a capability rather than a raw address.
If the provided capability is valid, $\ild$ and $\isd$ perform the operations
on its cursor.
Furthermore,
if the data word transferred is a capability that is linear, the original copy,
be it in a register (as in $\isd$ and $\imov$) or memory (as in
$\ild$), will be cleared to zero.
Formally, let $w$ be the data word, $\imov, \ild$ and $\isd$
set the content of the source location
(a register or a memory location) to $\fMoved(w)$:
\begin{equation*}
    \fMoved(w) =
    \begin{cases}
        0 & w = \capfd \land t \in \fLinearTypes \\
    w & \text{otherwise},
    \end{cases}
\end{equation*}
where (parameters are omitted here for
brevity)
\begin{equation*}
\fLinearTypes = \{\tLin, \tRev, \tUninit, \tSealed, \tSealedRet\}.
\end{equation*}

\subsection{Capability Revocation} 
\label{subsec:revocation}

\paragraph{Revocation}
We model the $\irevoke$ instruction 
using the \emph{revocation tree}.
The root of the revocation tree is a special node $\nroot$.
Each \emph{valid} capability $c$, 
regardless of its type, maps to a node with the index $c.n$
(we will
use an index to refer to a node for simplicity)
in the
revocation tree, whereas each revoked capability maps to
one outside the revocation tree (i.e., disconnected from $\nroot$).
Using $\irevoke$ on a revocation capability $r$
reparents all the children of $r.n$ to
$\nnull$, cutting the subtree off the revocation tree and thus
invalidating the nodes inside.
Meanwhile, the type of $r$ is changed to $\tLin$
(linear)
if the only capabilities that map to nodes in the subtree are
non-linear, or $\tUninit$ (uninitialized) otherwise.

\paragraph{Creation of revocation capabilities}
The $\imrev$ (``mint revocation'') instruction creates
a revocation capability from a linear capability $c$.
The resulting 
revocation capability $r$ receives the same
region bound and access permission set in $c$.
Meanwhile, a new node is created 
in the revocation tree
for $r$ between $c.n$ and its
parent.
Since $c.n$ is in the subtree of $r.n$, $c$ is revoked
in the process of $\irevoke\ r$.
This process naturally captures the hierarchical
strengths of revocation capabilities.
Consider the example where $r_1$ and $r_2$ are both created
using $\imrev$ on the same linear capability $c$, and $r_2$ is
created after $r_1$.
In this case, $r_1.n$ will be the parent of $r_2.n$, which is
in turn the parent of $c.n$.
Using $\irevoke$ on $r_1$ will therefore revoke both $r_2$ and
$c$, whereas using it on $r_2$ will only revoke $c$.

\paragraph{Uninitialized capabilities}
Uninitialized capabilities always grant 
only write memory access regardless of the permissions recorded in
them.
An uninitialized capability newly generated by
$\irevoke$ always has its cursor set to its base
address.
%
%
Every subsequent write made with the uninitialized
capability increments its cursor by one word position.
The cursor effectively marks the boundary of
the already initialized part of the memory
region.
\codename{} provides $\iinit$ to convert an
uninitialized capability whose cursor 
has reached its end address (and
thus fully initialized)
to a linear capability which inherits
both its memory region and access permissions.

\paragraph{Capability dropping}
The $\idrop$ instruction directly invalidates a
given capability.
By invoking $\idrop$ on a linear capability,
a domain effectively informs the
\codename{} implementation that it is not interested in the
memory region any more.
This can prevent $\irevoke$ from 
producing an unintended uninitialized capability.
Invoking $\idrop$ on a capability $c$ (which is
not non-linear) removes the node $c.n$ from the
revocation tree.
The children of $c.n$, if any, will be adopted by its parent.

\subsection{Capability Modification}

Instructions $\itighten, \ishrink$ and $\isplit$
modify a given capability without changing its
type.
The $\itighten$ instruction changes the memory access
permissions to a more restrictive subset.
The $\ishrink$ instruction sets the region bound to a
specified bound fully covered by the original one.

The $\isplit$ instruction splits a capability $c$ into 
two at the specified address $s$.
%
%
Let $b$ and $e$ be the base and end addresses of
the given capability with $b < s < e$,
then the two
resulting capabilities $c_1, c_2$ will have base and end addresses
$b_1 = b, e_1 = s$ and $b_2 = s, e_2 = e$ respectively.
The original capability becomes unavailable.
Meanwhile, the node $c.n$ 
is also split in two,
both inheriting the original parent.

The $\idelin$ instruction converts a
linear capability into a non-linear capability
by simply changing its type to $\tNon$.
For all but sealed (including sealed-return) and
uninitialized capabilities, $\iscc$
sets the cursor to a give address.
Another instruction, $\ilcc$, returns the cursor of a given
capability.

\subsection{Domain Switching}
%
%

\paragraph{Sealing} 
The $\iseal$ instruction converts a given
linear capability to a sealed capability.
The memory region associated with a sealed capability
contains the context of a security domain that is not currently
running.
An application can prepare desired contents using a linear capability,
and then $\iseal$ it into a sealed capability.
In this way, the application has effectively created a new domain with
a specific initial context.

A sealed capability does not grant direct memory access.
Rather, it needs to be \emph{unsealed} into the
register file of a physical thread either through the $\icall$
instruction or as the
result of an exception or interrupt, which effectively switches the
physical thread to the sealed domain.

\paragraph{Synchronous domain calls}
When a security domain $D$ holds a sealed capability $c_E$ for another
security domain $E$, $D$ can use $\icall$ on $c_E$
to switch the current physical thread to $E$.
This unseals $E$'s context from $c_E$ into the register file of the current
physical thread, while sealing $D$'s context (i.e., the current
content of the register file) to $c_E$'s associated region.
%
%
%
%
The $\icall$ instruction writes a sealed-return capability $c_r$
for the same region as $c_E$
to $E$'s $\rret$ register, so $E$ is able to return to $D$
later.
To facilitate communication between $D$ and $E$, the register $\rr_1$
is reserved for argument passing.
In other words, a second operand to $\icall$ is directly loaded into
$E$'s $\rr_1$ register.
To return to $D$, $E$ invokes $\ireturn$ on $c_r$.
The $\ireturn$ instruction is similar to $\icall$, 
except that it does not save the current context on the physical
thread.
Note that $c_E$ has been destroyed in $D$'s context during $\icall$.
$E$ can specify a value to $\ireturn$ to replace $c_E$ with when returning
to $D$.
A variant to $\ireturn$, $\iretseal$, replaces $c_E$ with $E$'s current context with
the $\rpc$ cursor switched to a specified value.
This is useful for allowing $D$ to invoke $E$ multiple times, each
time with a controlled and potentially different initial context.

\paragraph{Interrupt and exception handling}
\codename{} uses the same mechanism for $\iexcept$ as for $\icall$.
When an interrupt or exception occurs, the current
physical thread switches to a \emph{handler domain} defined
in the $\repc$ register of the current security domain.
An interrupt or exception is hence essentially an asynchronous $\icall$
on $\repc$.
%
%
%
%
%
%
The register receives the
following special treatments for its role in system management:

\emph{Pinned per-thread.} Except for interrupt or exception
handling, $\repc$ is excluded in the part of the
execution state replaced during domain switching.
The end result is that the $\repc$ value is per-thread
instead of per-domain;

\emph{Immutable.} Unless the current value is unset,
$\repc$ is immutable. In other words, $\repc$ is
fixed upon first write.

\section{Security Analysis}
\label{sec:security}

{Due to the space limit, we only briefly overview
the security proof.
Full details are available in Appendix~\ref{sec:absproofs}.}

We define an abstract model, \abscodename{}, and prove that
\codename{} refines it (main theorem).
The state of \abscodename{} is defined in Figure~\ref{fig:capstoneabs}.
Figure~\ref{fig:capstoneabsref} defines a refinement mapping between a
concrete state of \codename{} and an abstract state of \abscodename{}
with respect to a distinct domain $d$ which serves as the \emph{user
domain}, and a set of domains $D_\textsf{sub}$ which serves as the
\emph{subordinate environment}.

\begin{figure}[t]
    \centering
    {
\small
\begin{tabular}{rcl}
$\textit{dom}$ &$\coloneqq$& $\textsf{user $\mid$ sup $\mid$ sub}$\\
$\textit{mem}_{\textit{abs}}$ &$\coloneqq$& $\sAddr \mapsto (\sWord \mid \textsf{uninit})$\\
$\textit{range}$ &$\coloneqq$& $\{n \mid x \leq n < x+y \}$\\
$\textit{cap}_{\textit{abs}}$ &$\coloneqq$& $\textit{range}$\\
$\textit{tstate}$ &$\coloneqq$& $\textit{cap$_\textit{abs}$ set}$\\
$\textit{pstate}$ &$\coloneqq$& $(\textit{mem}_\textit{abs}, \textit{tstate}_\textsf{user}, \textit{tstate}_\textsf{sup},  \textit{tstate}_\textsf{sub})$\\
\end{tabular}}
\caption{\abscodename{} state.}
\label{fig:capstoneabs}
\end{figure}

\begin{figure}[t]
    \centering
    {
    \small
    \begin{tabular}{lll}
        \multicolumn{3}{l}{$\text{Refines}_\textit{(d,$D_\textsf{sub}$)}(\mstate,
        (\textit{mem}_\textit{abs}, \textit{tstate}_\textsf{user},
        \textit{tstate}_\textsf{sup},
        \textit{tstate}_\textsf{sub}))$} \\
        \quad & where:\ & $\textit{ind}_\textit{uninit} = \bigcup_{d
        \in \mathbb{N}}~\fworanges(\realm_w (\mstate, d))$ \\
        & & $\textit{mem}_\textit{abs} =
        \mstate.\textit{mem}[\textit{ind}_\textit{uninit} :=
        \textsf{uninit}]$ \\
        & &$\textit{tstate}_\textsf{user} =
        \text{ranges}(\xrealm(\mstate, d))$  \\
        & & $\textit{tstate}_\textsf{sub} = \bigcup_{d' \in
        D_\textsf{sub}}~\text{ranges}(\xrealm(\mstate, d'))$ \\
        & & $\textit{tstate}_\textsf{sup} = \bigcup_{d' \in (\mathbb{N} -
        D_\textsf{sub} - d)}~\text{ranges}(\xrealm(\mstate, d'))$ \\
    \end{tabular}}
\caption{\abscodename{} refinement mapping.}
\label{fig:capstoneabsref}
\end{figure}

\begin{sloppypar} We show that
    \codename{}'s use of uninitialized capabilities refines
    \abscodename{}'s use of \textsf{uninit} memory values to denote
    memory which cannot be accessed.
To this end, the line \mbox{$\textit{ind}_\textit{uninit} = \bigcup_{d
    \in \mathbb{N}}~\fworanges(\realm_w (\mstate, d))$} collects the
    ranges of all write-only (uninitialized) capabilities.
The line \mbox{$\textit{mem}_\textit{abs} =
\mstate.\textit{mem}[\textit{ind}_\textit{uninit} :=
\textsf{uninit}]$} indicates that the abstract memory is the same as
the concrete memory, except that indices for
uninitialized capabilities are set to \textsf{uninit}. \end{sloppypar}

In the abstract model, domains are represented as sets of abstract
capabilities, each being a simple range of
accessible addresses.
The definition $\xrealm(\mstate, d)$ is the \textit{exclusive realm}
of $d$, that is, the set of all concrete linear capabilities
exclusively accessible (transitively) in the domain $d$.
{The abstract state of the user domain (\mbox{$\textit{tstate}_\textsf{user}$})
is defined as the set of ranges corresponding to the capabilities in
the exclusive realm of $d$.}
{The abstract states of the subordinate and superordinate domains are
similarly defined.}

{We complete the proof by showing that the refinement mapping is preserved
by execution of the concrete model, and that steps in the concrete model can be
mapped to zero or more abstract actions in the abstract model.}

\section{Implementation}
\label{sec:impl-sketch}

We show that \codename{} can be implemented with acceptable
overhead.
Since a complete RTL implementation requires significant
engineering effort, we consider it as future work 
beyond the scope of this paper.
Instead, we present a sketch of a potential
implementation below, and evaluate it in
Section~\ref{sec:eval}.

\paragraph{Capabilities}
We represent each capability
as 128~bits in registers and memory as follows:
\vspace{-5pt}
\begin{center}
\includegraphics[width=0.9\linewidth]{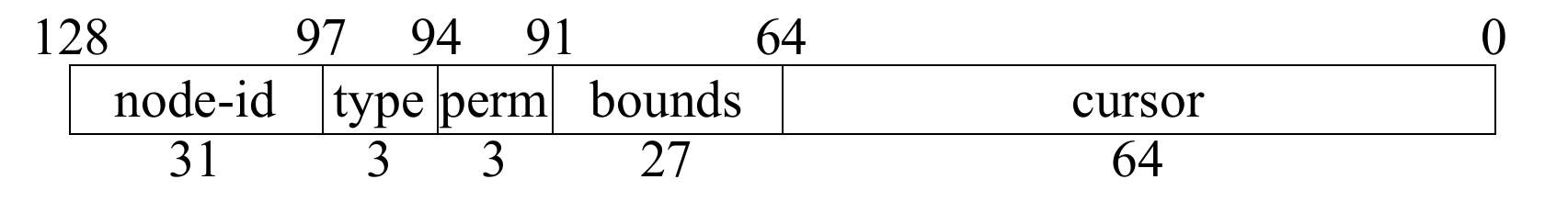}
\end{center}
\vspace{-10pt}
The \texttt{bounds} field encodes the capability address range
following the CHERI Concentrated scheme~\cite{cheri} which compresses
such information into 27 bits.
The \texttt{type} and \texttt{perm} fields indicate the type 
and associated permissions of each capability, and
the identifier of the associated revocation node of a capability is
recorded in \texttt{node-id}.
Each general-purpose register is extended to 16~bytes to allow it to
hold a capability.
To distinguish normal data from capabilities, we follow the
implementation of CHERI~\cite{cheri} to store a separate tag bit for each
register as well as every 16~bytes-aligned location in DRAM.
Existing work on implementing CHERI has shown that tag bits can be
maintained and queried efficiently~\cite{tagged-memory}.

\paragraph{Revocation tree}
On top of this, a \codename{} implementation also needs to record the
validity of each capability which might change due to revocations.
This concerns the maintenance of the revocation tree~(Section
\ref{subsec:revocation}).
Similarly to the tag bits, the nodes of the revocation tree are stored
in a DRAM region inaccessible to software.
Each revocation tree node is represented using the format below:
\vspace{-5pt}
\begin{center}
\includegraphics[width=0.9\linewidth]{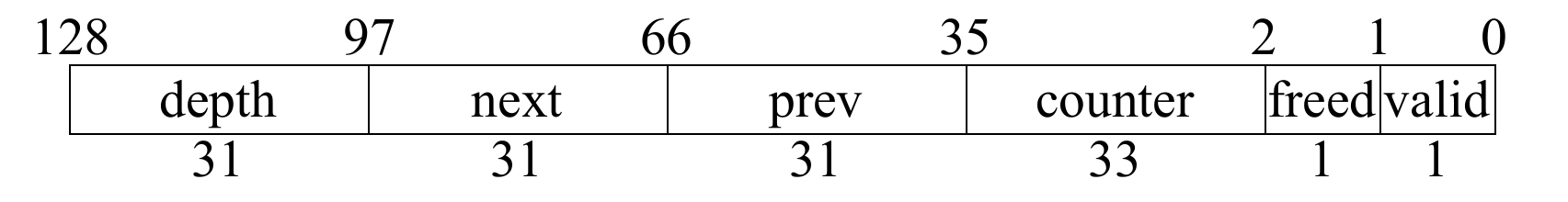}
\end{center}
\vspace{-10pt}
Whenever a capability is used in a memory access, the node associated
with it (\verb|node-id|) needs to be
retrieved from DRAM to query its validity.
To hide the latency of this query, we perform it in parallel to the
actual memory access.

\paragraph{Revocation}
A revocation operation 
involves traversing a subtree of a given node (the one associated
with the revocation capability) and invalidates each node within the
subtree.
Invalidated nodes are removed from the revocation tree so each
node can be visited and invalidated at most once.
This entails a constant amortized overhead of the revocation operation.
To facilitate subtree traversals, we maintain the revocation
tree as a doubly-linked list of nodes in the depth-first
order, with the depth recorded in each node.
Each subtree corresponds to a contiguous range within the linked list.
We unlink nodes from the linked list to remove them from the
revocation tree.

\paragraph{Deallocation of revocation nodes}
We cannot make a revocation node available for
allocation again immediately after invalidating it, as capabilities
that reference it may still exist.
However, we do need to free it at some point
because the DRAM region for storing revocation nodes is limited in
size.
We have two potential solutions to this issue.
One is a garbage collection mechanism based on memory sweeping:
Whenever free nodes run out,
we scan the whole memory to discover and free such nodes
that are not referenced by any capability.
The other option is to include a reference count in each node and free
a node when its reference count is zero.
We adopt the latter option, as we expect the former to introduce large
latencies in unpredictable locations.
This can be avoided with memory sweeps in parallel to the
pipeline execution, which, however, can be tricky to implement.
In comparison, reference counting would require updating
the counter when a capability is created or overwritten
but we expect the implementation
to be straightforward and the overhead acceptable.
We free a revocation node by adding it to a \texttt{free-nodes} linked
list.

\section{Evaluation}
\label{sec:eval}

We aim to answer the following question:
\emph{How does the performance of a \codename{} implementation compare
with that of a traditional platform?}
We use the gem5 simulator~\cite{gem5} to model the most performance-relevant
aspects of the implementation described in
Section~\ref{sec:impl-sketch}, namely the operations on the revocation
tree, including allocations, revocations, queries, and reference count
updates.
{Other parts such as bound and permission checking and
tagged memory are either trivial or already examined in previous work
in terms of implementation and performance impact~\cite{tagged-memory,pump,pump2,tagged-arch-guide}.}

\begin{figure}[t]
    \centering
    \includegraphics[width=.7\linewidth]{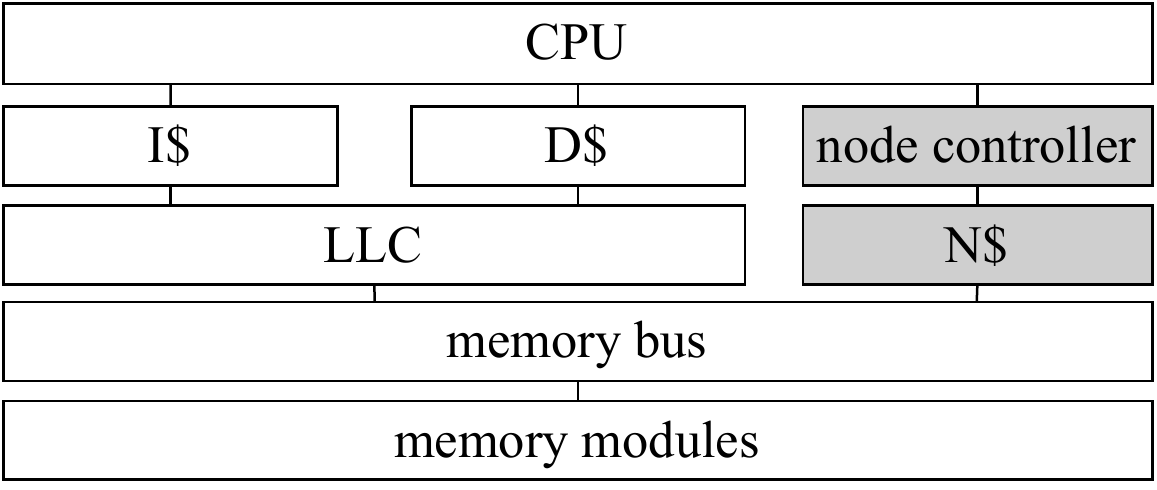}
    \caption{Overview of the \codename{} model implemented in gem5.
    The shaded components are added by us.}
    \label{fig:impl-overview}
\end{figure}

\paragraph{Setup}
In the absence of applications written for \codename{},
we map runtime behaviours of existing RISC-V applications
to the expected corresponding events
in their \codename{} ports. 
The details of this mapping are shown in Table~\ref{tab:mapping}.
As shown in
Figure~\ref{fig:impl-overview}, compared to a traditional system,
our gem5-based \codename{} model has no MMU but incorporates 
a node controller and a node cache for
revocation node storage.
{While out-of-order CPU models are more accurate for
modern mainstream high-performance systems,
we choose an in-order core 
for its lower complexity which 
is conducive to a first-step evaluation.
Evaluating on out-of-order models is future work.
}
We use a clock frequency of 1~GHz, a 2-way set associative L1 instruction (16~kB) and data cache (64~kB), and an 8-way set associative last level cache of 256~kB. The node cache (N\$) is an 8~kB 2-way set associative cache with a 32~GB 1600~MHz DDR3 DRAM.

\begin{table}[t]
    \centering
    \caption{Mappings from existing RISC-V application behaviours to events in \codename{} used in our evaluation.}
    \label{tab:mapping}
    \resizebox{\linewidth}{!}{
    \small
    \begin{tabular}{ll}
    \hline
        {\textbf{Behaviour in applications}} &
        \textbf{Event in \codename{}} \\
    \hline
        {\texttt{malloc}} & a new linear capability \\
        \hline
        {\texttt{free}} & revoking on a revocation capability \\
        \hline
        {overwriting an address} & destroying a nonlinear capability
        \\ \hline
        {producing an address} & creating
        a nonlinear capability
        \\
    \hline
    \end{tabular}}
\end{table}

\paragraph{Benchmarks}
We use the SPEC CPU 2017 intspeed benchmark suite~\cite{spec17}, ref inputs.
Instead of a full detailed simulation, which would take months to years to complete,
The MMU is removed from both simulations.

\paragraph{Results}
As shown in Table~\ref{tab:eval-results},
{
the workloads
vary widely in their use of revocation tree operations, 
ranging from no use (605.mcf\_s) to few to no allocations after initial setup (625.x264\_s, 631.deepsjeng\_s, and
657.xz\_s), to significant use of all operations. 
Correspondingly,}
the overhead varies from $0$ to $50\%$.
As expected, the overhead roughly correlates with the number of misses in the
node cache, as each of them involves accessing the DRAM.
The results also show that reference count updates are often the
dominating revocation tree operations in terms of frequency, and their
frequency strongly correlates with the displayed overhead.
We believe that this is partly caused by the inability to
hide the latency when a reference count update
results from a non-load/store instruction (e.g.,
move or pointer arithmetic).
In an actual \codename{} program, we expect such cases to be
considerably less frequent because of the ubiquity of linear
capabilities (for example, moving a linear capability
between registers does not change the reference count).
This also points to potential future work of exploring such optimizations
as delayed updates to better hide
the latency and improve the performance.

\begin{table*}[t]
    \centering
    \caption{Evaluation results on SPEC CPU 2017 intspeed, collected
    after 10~billion instructions of fast-forwarding.}
    \label{tab:eval-results}
    \resizebox{\linewidth}{!}{
        \small\begin{tabular}{lrrrrrrrrrr}
        \hline
        \multirow{2}{*}{\textbf{Workload}} &
        \multicolumn{2}{c}{\textbf{Runtime (seconds)}} & 
        \multirow{2}{*}{
        \textbf{Overhead
        (\%)}} &
        \multicolumn{3}{c}{\textbf{\codename{} node cache}} &
        \multicolumn{4}{c}{\textbf{\codename{} revocation tree
            operations}}
        \\
        & \multicolumn{1}{c}{\textbf{\codename{}}} & \multicolumn{1}{c}{\textbf{Baseline}} & &
        \multicolumn{1}{c}{\textbf{Misses}} & \multicolumn{1}{c}{\textbf{Hits}} & \multicolumn{1}{c}{\textbf{Miss rate (\%)}} &
        \multicolumn{1}{c}{\textbf{\#Allocation}} & \multicolumn{1}{c}{\textbf{\#Query}} &
        \multicolumn{1}{c}{\textbf{\#RC-update}} & \multicolumn{1}{c}{\textbf{\#Revocation}}
        \\ \hline 600.perlbench\_s/0 & 4.893 & 3.632 & 34.710 & 555427 & 615721998 & 0.090 & 22402 & 96213853 & 259932183 & 18754 \\
600.perlbench\_s/1 & 4.960 & 3.882 & 27.781 & 415766 & 529275069 & 0.078 & 59565 & 106558849 & 211290520 & 58279 \\
600.perlbench\_s/2 & 5.157 & 3.841 & 34.262 & 997293 & 625836524 & 0.159 & 1293 & 95528533 & 265646893 & 1161 \\
602.gcc\_s/0 & 5.256 & 4.005 & 31.219 & 295175 & 616668184 & 0.048 & 139829 & 98004533 & 258869083 & 139020 \\
602.gcc\_s/1 & 5.259 & 4.007 & 31.263 & 305182 & 617601867 & 0.049 & 139870 & 98162127 & 259262035 & 139032 \\
602.gcc\_s/2 & 5.260 & 4.007 & 31.257 & 325157 & 617383805 & 0.053 & 139814 & 98120826 & 259183800 & 139011 \\
605.mcf\_s/0 & 5.467 & 5.467 & 0.000 & 0 & 0 & 0.000 & 0 & 0 & 0 & 0 \\
620.omnetpp\_s/0 & 7.947 & 5.267 & 50.870 & 12058732 & 902225434 & 1.319 & 430165 & 172949008 & 368670213 & 379251 \\
623.xalancbmk\_s/0 & 9.017 & 6.202 & 45.387 & 16891678 & 851610538 & 1.945 & 112500 & 273464398 & 297073283 & 80999 \\
625.x264\_s/0 & 4.434 & 3.679 & 20.516 & 95 & 377394651 & 0.000 & 38 & 175363916 & 101015339 & 0 \\
625.x264\_s/1 & 4.511 & 3.765 & 19.819 & 0 & 373093048 & 0.000 & 0 & 183377516 & 94857766 & 0 \\
625.x264\_s/2 & 4.078 & 3.459 & 17.902 & 33 & 309551709 & 0.000 & 116 & 146974430 & 81288420 & 2 \\
631.deepsjeng\_s/0 & 3.363 & 3.344 & 0.565 & 0 & 9452119 & 0.000 & 0 & 4280475 & 2585822 & 0 \\
641.leela\_s/0 & 3.386 & 3.105 & 9.072 & 1764704 & 87073176 & 1.986 & 23432 & 33280704 & 27675933 & 19844 \\
648.exchange2\_s/0 & 3.646 & 3.638 & 0.222 & 0 & 4029952 & 0.000 & 22078 & 1909208 & 972060 & 22078 \\
657.xz\_s/0 & 3.011 & 2.899 & 3.846 & 0 & 33256028 & 0.000 & 0 & 24748672 & 4253678 & 0 \\
657.xz\_s/1 & 3.522 & 3.078 & 14.414 & 0 & 221848643 & 0.000 & 0 & 73389965 & 74229339 & 0 \\
\hline

    \end{tabular}}
\end{table*}

\section{Case Studies}
\label{sec:app}

\begin{lstfloat}
\begin{minipage}[t]{.47\linewidth}
\begin{Highlighting}[]
\KeywordTok{struct}\NormalTok{ capstone\_runtime \{}
 \DataTypeTok{void}\NormalTok{* malloc;}
 \DataTypeTok{void}\NormalTok{* free;}
 \DataTypeTok{void}\NormalTok{* thread\_start;}
 \DataTypeTok{void}\NormalTok{* thread\_create;}
 \DataTypeTok{void}\NormalTok{* join\_all;}
 \DataTypeTok{void}\NormalTok{* enclave\_create;}
 \DataTypeTok{void}\NormalTok{* enclave\_enter;}
 \DataTypeTok{void}\NormalTok{* enclave\_destroy;}
\NormalTok{\};}
\end{Highlighting}
\end{minipage}
\begin{minipage}[t]{.47\linewidth}
\begin{Highlighting}[]
\KeywordTok{struct}\NormalTok{ mem\_region \{}
 \KeywordTok{struct}\NormalTok{ mem\_region *left,}
  \NormalTok{*right;}
 \DataTypeTok{int}\NormalTok{ size, leaf, free;}
 \DataTypeTok{void}\NormalTok{* mem;}
\NormalTok{\};}
\KeywordTok{struct}\NormalTok{ malloc\_state \{}
 \KeywordTok{struct}\NormalTok{ mem\_region* heap;}
 \DataTypeTok{int}\NormalTok{ alloc\_n;}
\NormalTok{\};}
\end{Highlighting}
\end{minipage}
\vspace{-2pt}
\caption{{Left: Data structure that exposes runtime interfaces. Right:
Data structures in the memory allocator.}}
\label{lst:enclave-runtime}
\end{lstfloat}

\begin{table}[t]
    \centering
    \caption{{LoC of each component of the prototype implementation: \emulatorname{},
    \compilername{}, and \libraryname{}.}}
    \label{tab:prototype-loc}
    {
    \small
    \begin{tabular}{cccc}
    \hline
       \textbf{\codename{}-{}}  & \textbf{Emu} & \textbf{CC} & \textbf{Lib} \\ 
       \hline
       \textbf{LoC} & 1081 & 2319 & 529 \\ \hline
    \end{tabular}
    }
\end{table}

The expressiveness of \codename{} enables
the memory isolation models 
discussed in Section~\ref{subsec:examples}.
To demonstrate this, 
we have implemented a \emph{functional} prototype of \codename{} in the form
of an ISA emulator (\emulatorname{})
and a simple compiler (\compilername{}), and, on top of them,
a runtime library (\libraryname{})
that encapsulates \emph{runnable} implementations of 
different memory isolation models.
{Table~\ref{tab:prototype-loc} summarizes the lines of code in each component.}

Unlike our gem5 model (see Sections~\ref{sec:impl-sketch}
and \ref{sec:eval}), \emulatorname{} is intended purely for
exploring the expressiveness of the \codename{} \emph{interface}.
Therefore, we keep its implementation high-level and straightforward. 
\compilername{} compiles a
C-like language into \codename{}.
\libraryname{}
exposes interfaces to memory isolation models
through sealed capabilities inside
a \verb|capstone_runtime| object {(Listing~\ref{lst:enclave-runtime}, left)}.
{We have open-sourced those tools and our case study
implementations (see \hyperref[sec:availability]{Availability}).}
{
We summarize the lines of code (LoC) of each
case study in Table~\ref{tab:case-study-loc}.}

\begin{table}[t]
    \centering
    \caption{{LoC (input to \compilername{})
    of our case study implementations in \libraryname{}.
    Abbreviations:
    MA (memory allocator), TS (thread scheduler), Enc (enclaves).}}
    \label{tab:case-study-loc}
    {
    \small
    \begin{tabular}{ccccc}
    \hline
       \textbf{Case study} & \textbf{MA} & \textbf{TS} & \textbf{Encl} & \textbf{Rust} \\ \hline
        \textbf{LoC} & 128 & 242 & 75 & 5 \\ \hline
    \end{tabular}
    }
\end{table}

\paragraph{Trustless memory allocation}
Our heap memory allocator exposes
two interfaces to applications:
\verb|malloc| and \verb|free|.
The \verb|malloc| interface receives the size of the memory
region to be allocated, and returns a valid linear capability
if the allocation succeeds.
The \verb|free| interface receives a linear capability for a 
previously allocated memory region,
and makes it available for future allocations.
An application does not need to trust the memory allocator.
After obtaining a linear capability from \verb|malloc|, the
application is guaranteed exclusive access to the memory region.
The allocator may revoke the
capability, but cannot read the original memory content 
as long as the application holds its linear or
revocation capability.

The \verb|malloc_state|
object maintains the state of a memory allocator
{(Listing~\ref{lst:enclave-runtime}, right)}.
It contains a \verb|heap| capability that covers all memory
available for allocation.
Since the capability to \verb|malloc_state| is sealed inside 
\verb|malloc| and \verb|free| in \verb|capstone_runtime|,
an application can only access the allocator state
\emph{indirectly} through those well-defined interfaces.

In summary, our implementation guarantees that the memory
allocator can reclaim memory whenever it wants, but cannot
access any allocated region,
or read private data {in} a region after reclaiming it.
Trust between applications and the memory
allocator is thus unnecessary.
{
Since \codename{} does not include a centrally-managed
MMU,
mechanisms commonly relying on 
it
(e.g., swapping, copy-on-write) become
non-trivial.
Enabling them with capabilities 
is future work.
}

\paragraph{Trustless thread scheduling}
We implemented a thread scheduler that requires no trust from
applications.
The scheduler belongs to a different domain and has no access
to the linear capabilities held by the application domain (and, in
turn, the data they point to).
\codename{} ensures that
the application domain context is safely saved and
restored when an exception occurs and when the domain resumes
execution.

Part of the data involved
is critical in the sense that at most one thread can safely
manipulate it at any time.
To protect such data,
we encapsulate them in a
\verb|sched_critical_state| object, and include a \emph{linear} capability
to it inside the scheduler state \verb|sched_state|.
Before a thread accesses such data, it needs to 
load the linear capability into a register.
The linearity of the capability subsequently guarantees that no other thread
can access the data structure.

Besides preventing the thread scheduler from
accessing application data during context switches, our
implementation has several more security benefits.
Since the exception handler is defined by a normal
sealed capability, an application can attest to
the identity of the exception handler or thread scheduler (when sealed
capabilities are extended with cryptographic checksums).
This mitigates attacks that involve attacker-controlled
exception handlers or thread schedulers, such as 
Game of Threads~\cite{game-of-threads} and SmashEx~\cite{smashex}.
By safely storing
the domain context upon a context switch,
\codename{} also allows better control of domain re-entries,
as re-entries that accesses overlapping resources are impossible,
which improves the security of custom exception handling (e.g.,
in-enclave exception handling in Intel SGX~\cite{sgx,sgxexplained}).

\paragraph{Spatially-isolated enclaves}
We implemented a basic set of interfaces for a TEE with
spatially-isolated enclaves similar to Intel SGX~\cite{sgx,
sgxexplained} and Keystone~\cite{keystone}:
\verb|enclave_create|,
\verb|enclave_enter|, and \verb|enclave_destroy|.
The central data structures include \verb|enclave|,
which is available
to the software creating and using an enclave, 
and \verb|enclave_runtime|, which
is available to the enclave itself.

The \verb|enclave_create| interface creates a new enclave from
two input linear capabilities for its code and data respectively.
Both capabilities are then sealed together in a sealed
capability, alongside an enclave private stack and an
\verb|enclave_runtime| object.
Sealing protects the corresponding memory
regions from direct access outside the enclave itself,
similar to the enclave setup in
Intel SGX~\cite{sgx,sgxexplained}.
To facilitate data exchange between the host application and the enclave,
\verb|enclave_create| creates a shared memory region
between them, 
and its capability is
placed in both \verb|enclave_runtime|
and \verb|enclave|.
The \verb|enclave_enter| interface calls into the sealed capability
contained in a given \verb|enclave| structure,
effectively executing the enclave.
The \verb|enclave_destroy| interface
reclaims and frees the memory resources of an enclave
with the revocation capabilities inside \verb|enclave|.

{This case study focuses on memory isolation.
A complete
TEE platform usually also
includes a hardware root of trust,
memory encryption, and local and remote
attestation~\cite{sgx,sgxexplained,armtrustzone,keystone}.
We consider the hardware root of trust and memory
encryption as orthogonal to \codename{}.
For attestation, future work may explore
attaching measurements to
uninitialized capabilities
and extending them
upon each memory store.
The measurement is frozen
when the uninitialized capability is initialized,
and henceforth invalidated upon further stores.}

\paragraph{Nested enclaves}
In our spatially-isolated enclave implementation,
the domain creating an enclave can also be an enclave itself.
To enable nested enclaves, we only need to expose the enclave creation
interfaces to enclaves.
This is easily achieved by passing the \verb|capstone_runtime|
structure to each enclave inside the \verb|enclave_runtime| structure.
The nesting structure can be extended
indefinitely during runtime on demand.
Each enclave can be sure that a memory region shared with a child
enclave is only accessible to this same child enclave or those nested
inside it and can be reclaimed at any time.

\begin{lstfloat}
\begin{Highlighting}[]
\DataTypeTok{void}\NormalTok{* shared\_mem = capstone\_runtime{-}\textgreater{}malloc(}\DecValTok{128}\NormalTok{);}
\NormalTok{shared\_mem[}\DecValTok{0}\NormalTok{] = }\DecValTok{42}\NormalTok{; }\CommentTok{// just some dummy data}
\DataTypeTok{void}\NormalTok{* shared\_rev = mrev(shared\_mem);}
\DataTypeTok{void}\NormalTok{* shared\_rev\_shared = mrev(shared\_mem);}
\NormalTok{drop(shared\_mem); }\CommentTok{// drop to share}
\NormalTok{runtime{-}\textgreater{}heap[}\DecValTok{1}\NormalTok{] = shared\_rev;}
\DataTypeTok{void}\NormalTok{* emissary = }
  \NormalTok{capstone\_runtime{-}\textgreater{}}\NormalTok{malloc(CAPSTONE\_SEALED\_SIZE);}
\NormalTok{scco(emissary\_code, }\DecValTok{0}\NormalTok{);}
\NormalTok{emissary[CAPSTONE\_OFFSET\_PC] =}\NormalTok{ emissary\_code;}
\NormalTok{emissary[CAPSTONE\_OFFSET\_EPC] = }\DecValTok{0}\NormalTok{;}
\NormalTok{emissary[CAPSTONE\_OFFSET\_DEDICATED\_STACK] = }\DecValTok{0}\NormalTok{;}
\NormalTok{emissary[CAPSTONE\_OFFSET\_METAPARAM] = }\NormalTok{shared\_rev\_shared;}
\NormalTok{seal(emissary);} \NormalTok{runtime{-}\textgreater{}shared[}\DecValTok{0}\NormalTok{] = emissary;}
\end{Highlighting}
\vspace{-2pt}
\caption{{Preparing a shared memory region.}}
\label{lst:enclave-shmem-prep}
\end{lstfloat}

\begin{lstfloat}
\begin{minipage}[t]{.45\linewidth}
\begin{Highlighting}[]
\DataTypeTok{void}\NormalTok{* d = }
  \NormalTok{runtime{-}\textgreater{}shared[}\DecValTok{0}\NormalTok{];}
\DataTypeTok{void}\NormalTok{* shared\_mem = d();}
\NormalTok{revoke(shared\_mem);}
\end{Highlighting}
\end{minipage}
\begin{minipage}[t]{.45\linewidth}
\begin{Highlighting}[]
\NormalTok{CAPSTONE\_ATTR\_HAS\_METAPARAM }
\DataTypeTok{void}\NormalTok{* setup\_shared() \{}
 \DataTypeTok{void}\NormalTok{* d = CAPSTONE\_METAPARAM;}
 \CommentTok{// check measurement of ret}
 \ControlFlowTok{return}\NormalTok{ d;}
\NormalTok{\}}
\end{Highlighting}
\end{minipage}
\vspace{-2pt}
\caption{{Left: accessing a shared memory region. Right: domain that performs authentication and returns revocation capabilities for shared memory regions.}}
\label{lst:enclave-shmem-acc}
\end{lstfloat}

\paragraph{Temporally-isolated enclaves}
Since \codename{} does not rely on identity-based access
control, an enclave $D$ cannot directly share a memory
region exclusively with another enclave $E$, unless $E$ is created by
$D$ or $D$ can access the sealed capability of $E$ through other
means.
In general, $D$ needs to pass a capability to $E$ and $E$ alone.
To achieve this on \codename{}, $D$ can create a domain $C$ specially for
communicating with $E$ and then pass $C$'s sealed capability
to $E$ {(Listing~\ref{lst:enclave-shmem-prep})}.
The host then marshalls $C$'s sealed capability
to $E$, which obtains access to the shared memory region by
invoking $C$ and then performing revocation with the returned revocation
capability {(Listing~\ref{lst:enclave-shmem-acc}, left)}.
$C$ can then perform authentication, e.g., by examining
the measurement of the sealed-return capability in $\rret$, to make sure
that it is invoked by $E$ before provisioning a revocation
capability for the shared memory region {(Listing~\ref{lst:enclave-shmem-acc}, right)}.
{Hardware-generated cryptographic checksums are beyond the scope
of this paper.}

Note that 
$D$ can limit $E$'s access
permissions to the memory region through the permissions in the
revocation capability passed to $E$.
In addition, since $D$ holds another revocation capability
created before the one passed to $E$, it can revoke
the delegated access at any time.
To establish a non-exclusive
shared memory region
with $E$, $D$ may have $C$ pass to $E$ a
non-linear capability instead of a revocation capability.
By passing linear capabilities back and forth through the
non-exclusive shared memory between the two enclaves, they can
take turns to have exclusive access to other memory regions in
multiple rounds with the non-exclusive region as a trampoline.

Our implementation prevents 
unintended enclaves from accessing a temporarily shared memory
region.
Through revocation capabilities, it also allows
an accessor to obtain exclusive access.
Moreover, the owner enclave of a memory region can
limit what each accessor can do to it.
%

%
%
%

%

\begin{table}[t]
    \centering
    \caption{{Rust-like abstraction on \codename{}.}}
    \label{table:rust-abstraction}
    {
        \footnotesize
    \begin{tabular}{p{.26\linewidth}lp{.35\linewidth}}
        \hline
        \textbf{Op.}& \textbf{Rust} & \textbf{\codename{}} \\
        \hline
        Move & {\texttt{let a = b;}} &
        {\texttt{mov ra rb}} \\ \hline
        Immutable borrow & {\footnotesize\texttt{let a = \&b;}} &
        {\texttt{mrev rr rb; delin rb; li r0 0;
        tighten rb r0; mov ra rb; (use ra) revoke rr; mov rb rr}} \\ \hline
        Mutable borrow & {\footnotesize\texttt{let a = \&mut b;}} &
        {\texttt{mrev rr rb; mov ra rb; (use ra) revoke rr; mov rb rr}} \\
        \hline
    \end{tabular}
    }
\end{table}

\paragraph{Rust-like memory restrictions}
\codename{}
can enforce
Rust-like memory restrictions across security domains
at runtime
without assuming trusted software components.
{Table~\ref{table:rust-abstraction} summarizes the mapping from 
Rust operations to the corresponding
\codename{} primitives.}
{Owner references in Rust are directly mapped to linear capabilities, as they are similarly alias-free and non-duplicable.}
However, \codename{} has no direct equivalent to the mutable borrowed
reference.
Instead, 
we
pass the linear capability itself for mutable borrowing, and utilize
the revocation capability to ensure its return (in Rust, the owner reference
becomes usable again after the lifetime of the borrowed references
ends).
Immutable borrowing is supported
through read-only non-linear capabilities created by delinearizing the
linear capability and then tightening the permissions to read-only,
which can then be shared in arbitrarily
many copies, matching the behaviours of immutable borrowed
references in Rust.
Again, the domain uses revocation capabilities to ensure that its
exclusive access (owner reference) can be reclaimed.

\section{Related Work}

\paragraph{Architectural capabilities}
Early computer architectures with capability-based memory
addressing can be traced back to the early
1980s, but failed to see widespread adoption due to significant performance
overhead~\cite{system38}.
M-machine~\cite{mmachine} improved the performance through tagged
memory words and a shared address space across all protection domains.
HardBound~\cite{hardbound} proposed a limited form of architectural
capabilities without unforgeability to improve the
performance of bounds-checking in C programs.
More recently, CHERI~\cite{cheri} follows the tagged memory design of
M-machine with improved memory region granularity
and compatibility with traditional page-based memory protections.
Unlike \codename{}, all those designs assume a
trusted OS kernel, and
are unable to express exclusive access guarantees or
hierarchical capability revocation.
{Instead of relying on capability
metadata,
\cccname{} uses pointer encryption and
memory encryption to prevent secret leakage and predictable
memory tampering~\cite{crypto-cap}, which helps reduce its performance
overhead.
However, this trades off its flexibility in expressing
more sophisticated rules such
as those associated with different capability types.%
}
Capability-based security has also seen adoption in 
software designs, including
OS kernels~\cite{keykos, barrelfish,sel4,capsicum}, programming
languages~\cite{pony, e-lang}, and web services~\cite{oauth2}.
Such designs deal with higher-level notions of resources rather than
memory.

\paragraph{Linear capabilities}
Naden et al.\ proposed a type system with ``unique permissions'', a
concept similar to linear capabilities, to achieve efficient flexible
borrowing~\cite{borrowing-permissions}.
This is different from \codename{} which provides linear capabilities
at the lower architectural level, and enforces restrictions during
runtime.
StkTokens~\cite{stktokens} is a calling convention that utilizes
architectural linear capabilities to provide control flow
integrity in the context of software fault isolation.
StkTokens
is focused on a specific memory model,
whereas \codename{} intends to support multiple models at the same
time.
Moreover, 
StkTokens does not discuss scenarios with
asynchronous exceptions or when
untrusted software refuses to relinquish a linear capability.
It is also unclear how linear and non-linear capabilities interact.
Van Strydonck et al.\ proposed capturing
spatial separation logic predicates during runtime through compiling
verified C code into a low-level language with linear
capabilities~\cite{linear-cap-separation-logic}.
We consider their work as orthogonal to ours,
as \codename{} is focused exclusively on low-level interfaces.
The CHERI ISA document~\cite{cheri} briefly discusses an incomplete
linear capability design as an experimental feature to replace
garbage collection.
It is unclear from the document what interfaces related to linear
capabilities are available.
Moreover, CHERI relies on a trusted OS kernel to manage linear capabilities.
They propose that the OS kernel be allowed to violate
linearity to this purpose.
This fundamentally contrasts the goal of \codename{}.

\paragraph{Uninitialized capabilities}
Georges et al.\ introduced the notion of uninitialized
capability as a mechanism to improve the performance of capability
revocation~\cite{uninitialised-cap}.
Converting a normal capability into an uninitialized capability allows
invalidating the capabilities that reside in a large memory region
without requiring a scan through it.
The application can hence gain the guarantee that a memory region does
not contain any capability with a small constant overhead.
Unlike their work, 
\codename{} generalizes
uninitialized capabilities to the generic role of preventing secret leakage,
where the secrets
include, but are no more limited to, capabilities.

\paragraph{Linear and uniqueness type systems}
Some high-level programming languages have adopted linear type
systems~\cite{linear-logic-use, linear-unique, rust, haskell2010,
pony} or uniqueness type systems~\cite{linear-unique,
clean-lang-uniqueness}.
Linear types require that values be used exactly once in
the future,
whereas uniqueness types require that the values have never
been duplicated in the past.
They both allow conversions in single directions:
from
an unrestricted (i.e., non-linear) type into a linear
type, 
or from a uniqueness type into an unrestricted
type~\cite{linear-unique, linear-logic-use, clean-lang-uniqueness}.
In contrast,
by providing a mechanism for revocation, \codename{} enables
conversions (i.e., linearization and delinearization) in both
directions.
Unlike high-level programming languages,
\codename{} is intended as a low-level interface that enforces similar
restrictions during runtime.

{
\paragraph{Capability revocation}
Prior work on CHERI capability revocation
adopts different semantics to the term ``revocation''
than \codename{} does.
Whereas revocation in \codename{} immediately invalidates
a set of capabilities defined by the revocation hierarchy to
reclaim a linear capability,
CHERIvoke~\cite{cherivoke} and Cornucopia~\cite{cornucopia}
lazily invalidate capabilities through memory sweeping to ensure
that objects do not have stale capabilities associated with them
when they are reallocated (i.e., no use-after-reallocation).
While \codename{} is stricter in requiring 
invalidations to immediately take effect, its more constrained
capability provisioning operations help simplify tracking of
capability derivations and facilitate
capability revocation.

\paragraph{Tagged architectures}
Similar to CHERI~\cite{cheri-cap-model,tagged-memory},
\codename{} is a tagged architecture~\cite{tagged-arch-guide}
as it uses hardware-maintained word-granular 
metadata to distinguish capabilities
and enforce corresponding memory access isolation policies.
Other prior tagged architectures mostly focus on accelerating
specific security policies such as
control flow integrity (CFI) and spatial memory safety~\cite{arm-mte},
whereas \codename{} aims at providing a novel capability-based
memory access model.
Designs for general-purpose software-defined tag computations
have been proposed~\cite{pump,pump2},
but as the tag computations in \codename{} are concrete and sufficiently simple
to implement in hardware, have a huge input space (e.g., address ranges),
and are sometimes not purely functional (e.g.,
revocation nodes), those designs are ill-suited for supporting \codename{}.
}

\section{Conclusions}

We have proposed \codename{}, a new capability-based architectural
design that provides the flexibility to support multiple memory
isolation models without assuming trusted software components.
We pointed out that existing designs are
insufficient to achieve such a goal and described the additions
needed to overcome those limitations, 
specifically through a careful design with
linear and revocation capabilities.
Our evaluation results suggest that \codename{} can be implemented
with acceptable overhead.
As future work, we plan to explore hardware implementations of
\codename{} and to experiment with a wider range of use cases.

{
\section*{Acknowledgments}
We thank the anonymous reviewers,
Shweta Shinde, and Bo Wang for their suggestions on earlier drafts
of this paper.
This research is supported by the Ministry of Education, Singapore, under its Academic Research Fund Tier~1, T1~251RES2023 \mbox{(A-0008125-00-00)}.
Trevor E. Carlson is supported by a research grant from Huawei.
Conrad Watt is supported by a Research Fellowship from Peterhouse, University of Cambridge.
Any opinions,
findings, and conclusions or recommendations expressed in this
material are those of the authors only.

\section*{Availability}
\label{sec:availability}
The code produced in this work
is publicly available at \url{https://github.com/jasonyu1996/capstone}.
}

\bibliographystyle{plain}
\bibliography{main}

\begin{thebibliography}{10}

\bibitem{armtrustzone}
Arm trustzone technology.
\newblock \url{https://developer.arm.com/ip-products/security-ip/trustzone}.

\bibitem{js-cap-leak}
Adam Barth, Joel Weinberger, and Dawn Song.
\newblock Cross-origin javascript capability leaks: Detection, exploitation,
  and defense.
\newblock In {\em Proceedings of the 18th Conference on USENIX Security
  Symposium}, SSYM'09, page 187–198, USA, 2009. USENIX Association.

\bibitem{barrelfish}
Andrew Baumann, Paul Barham, Pierre-Evariste Dagand, Tim Harris, Rebecca
  Isaacs, Simon Peter, Timothy Roscoe, Adrian Sch\"{u}pbach, and Akhilesh
  Singhania.
\newblock The multikernel: A new os architecture for scalable multicore
  systems.
\newblock In {\em Proceedings of the ACM SIGOPS 22nd Symposium on Operating
  Systems Principles}, SOSP '09, page 29–44, New York, NY, USA, 2009.
  Association for Computing Machinery.

\bibitem{multics}
A.~Bensoussan, C.~T. Clingen, and R.~C. Daley.
\newblock The multics virtual memory: Concepts and design.
\newblock {\em Commun. ACM}, 15(5):308–318, may 1972.

\bibitem{burroughs}
H~Bingham.
\newblock Access controls in burroughs large systems.
\newblock {\em Privacy and Security in Computer Systems}, pages 42--45, 1974.

\bibitem{gem5}
Nathan Binkert, Bradford Beckmann, Gabriel Black, Steven~K. Reinhardt, Ali
  Saidi, Arkaprava Basu, Joel Hestness, Derek~R. Hower, Tushar Krishna, Somayeh
  Sardashti, Rathijit Sen, Korey Sewell, Muhammad Shoaib, Nilay Vaish, Mark~D.
  Hill, and David~A. Wood.
\newblock The gem5 simulator.
\newblock {\em SIGARCH Comput. Archit. News}, 39(2):1–7, aug 2011.

\bibitem{spec17}
James Bucek, Klaus-Dieter Lange, and J\'{o}akim v.~Kistowski.
\newblock Spec cpu2017: Next-generation compute benchmark.
\newblock In {\em Companion of the 2018 ACM/SPEC International Conference on
  Performance Engineering}, ICPE '18, page 41–42, New York, NY, USA, 2018.
  Association for Computing Machinery.

\bibitem{mmachine}
Nicholas~P. Carter, Stephen~W. Keckler, and William~J. Dally.
\newblock Hardware support for fast capability-based addressing.
\newblock In Forest Baskett and Douglas~W. Clark, editors, {\em {ASPLOS-VI}
  Proceedings - Sixth International Conference on Architectural Support for
  Programming Languages and Operating Systems, San Jose, California, USA,
  October 4-7, 1994}, pages 319--327. {ACM} Press, 1994.

\bibitem{cheri-memory-safe-c}
David Chisnall, Colin Rothwell, Robert N.~M. Watson, Jonathan Woodruff, Munraj
  Vadera, Simon~W. Moore, Michael Roe, Brooks Davis, and Peter~G. Neumann.
\newblock Beyond the {PDP-11:} architectural support for a memory-safe {C}
  abstract machine.
\newblock In {\"{O}}zcan {\"{O}}zturk, Kemal Ebcioglu, and Sandhya Dwarkadas,
  editors, {\em Proceedings of the Twentieth International Conference on
  Architectural Support for Programming Languages and Operating Systems,
  {ASPLOS} 2015, Istanbul, Turkey, March 14-18, 2015}, pages 117--130. {ACM},
  2015.

\bibitem{pony}
Sylvan Clebsch, Sophia Drossopoulou, Sebastian Blessing, and Andy McNeil.
\newblock Deny capabilities for safe, fast actors.
\newblock In {\em Proceedings of the 5th International Workshop on Programming
  Based on Actors, Agents, and Decentralized Control}, AGERE! 2015, page
  1–12, New York, NY, USA, 2015. Association for Computing Machinery.

\bibitem{sgxexplained}
Victor Costan and Srinivas Devadas.
\newblock Intel {SGX} explained.
\newblock {\em {IACR} Cryptol. ePrint Arch.}, 2016:86, 2016.

\bibitem{smashex}
Jinhua Cui, Jason~Zhijingcheng Yu, Shweta Shinde, Prateek Saxena, and Zhiping
  Cai.
\newblock Smashex: Smashing {SGX} enclaves using exceptions.
\newblock In Yongdae Kim, Jong Kim, Giovanni Vigna, and Elaine Shi, editors,
  {\em {CCS} '21: 2021 {ACM} {SIGSAC} Conference on Computer and Communications
  Security, Virtual Event, Republic of Korea, November 15 - 19, 2021}, pages
  779--793. {ACM}, 2021.

\bibitem{cheriabi}
Brooks Davis, Robert N.~M. Watson, Alexander Richardson, Peter~G. Neumann,
  Simon~W. Moore, John Baldwin, David Chisnall, Jessica Clarke,
  Nathaniel~Wesley Filardo, Khilan Gudka, Alexandre Joannou, Ben Laurie,
  A.~Theodore Markettos, J.~Edward Maste, Alfredo Mazzinghi, Edward~Tomasz
  Napierala, Robert~M. Norton, Michael Roe, Peter Sewell, Stacey Son, and
  Jonathan Woodruff.
\newblock Cheriabi: Enforcing valid pointer provenance and minimizing pointer
  privilege in the posix c run-time environment.
\newblock In {\em Proceedings of the Twenty-Fourth International Conference on
  Architectural Support for Programming Languages and Operating Systems},
  ASPLOS '19, page 379–393, New York, NY, USA, 2019. Association for
  Computing Machinery.

\bibitem{hardbound}
Joseph Devietti, Colin Blundell, Milo M.~K. Martin, and Steve Zdancewic.
\newblock Hardbound: architectural support for spatial safety of the c
  programming language.
\newblock In {\em ASPLOS 2008}, 2008.

\bibitem{pump2}
Udit Dhawan, Catalin Hritcu, Raphael Rubin, Nikos Vasilakis, Silviu Chiricescu,
  Jonathan~M. Smith, Thomas~F. Knight, Benjamin~C. Pierce, and Andre DeHon.
\newblock Architectural support for software-defined metadata processing.
\newblock {\em SIGARCH Comput. Archit. News}, 43(1):487–502, mar 2015.

\bibitem{pump}
Udit Dhawan, Nikos Vasilakis, Raphael Rubin, Silviu Chiricescu, Jonathan~M.
  Smith, Thomas~F. Knight, Benjamin~C. Pierce, and Andr\'{e} DeHon.
\newblock Pump: A programmable unit for metadata processing.
\newblock In {\em Proceedings of the Third Workshop on Hardware and
  Architectural Support for Security and Privacy}, HASP '14, New York, NY, USA,
  2014. Association for Computing Machinery.

\bibitem{penglai}
Erhu Feng, Xu~Lu, Dong Du, Bicheng Yang, Xueqiang Jiang, Yubin Xia, Binyu Zang,
  and Haibo Chen.
\newblock Scalable memory protection in the {PENGLAI} enclave.
\newblock In Angela~Demke Brown and Jay~R. Lorch, editors, {\em 15th {USENIX}
  Symposium on Operating Systems Design and Implementation, {OSDI} 2021, July
  14-16, 2021}, pages 275--294. {USENIX} Association, 2021.

\bibitem{uninitialised-cap}
A\"{\i}na~Linn Georges, Arma\"{e}l Gu\'{e}neau, Thomas Van~Strydonck, Amin
  Timany, Alix Trieu, Sander Huyghebaert, Dominique Devriese, and Lars
  Birkedal.
\newblock Efficient and provable local capability revocation using
  uninitialized capabilities.
\newblock {\em Proc. ACM Program. Lang.}, 5(POPL), jan 2021.

\bibitem{linear-logic}
Jean-Yves Girard.
\newblock Linear logic.
\newblock {\em Theoretical Computer Science}, 50(1):1--101, 1987.

\bibitem{oauth2}
Dick Hardt.
\newblock {The OAuth 2.0 Authorization Framework}.
\newblock RFC 6749, October 2012.

\bibitem{confused-deputy}
Norm Hardy.
\newblock The confused deputy: (or why capabilities might have been invented).
\newblock {\em SIGOPS Oper. Syst. Rev.}, 22(4):36–38, October 1988.

\bibitem{keykos}
Norman Hardy.
\newblock Keykos architecture.
\newblock {\em SIGOPS Oper. Syst. Rev.}, 19(4):8–25, oct 1985.

\bibitem{system38}
Merle~E. Houdek, Frank~G. Soltis, and Roy~L. Hoffman.
\newblock Ibm system/38 support for capability-based addressing.
\newblock In {\em Proceedings of the 8th Annual Symposium on Computer
  Architecture}, ISCA '81, page 341–348, Washington, DC, USA, 1981. IEEE
  Computer Society Press.

\bibitem{tagged-arch-guide}
Samuel Jero, Nathan Burow, Bryan Ward, Richard Skowyra, Roger Khazan, Howard
  Shrobe, and Hamed Okhravi.
\newblock Tag: Tagged architecture guide.
\newblock {\em ACM Comput. Surv.}, 55(6), dec 2022.

\bibitem{tagged-memory}
Alexandre Joannou, Jonathan Woodruff, Robert Kovacsics, Simon~W. Moore, Alex
  Bradbury, Hongyan Xia, Robert~N.M. Watson, David Chisnall, Michael Roe,
  Brooks Davis, Edward Napierala, John Baldwin, Khilan Gudka, Peter~G. Neumann,
  Alfredo Mazzinghi, Alex Richardson, Stacey Son, and A.~Theodore Markettos.
\newblock Efficient tagged memory.
\newblock In {\em 2017 IEEE International Conference on Computer Design
  (ICCD)}, pages 641--648, 2017.

\bibitem{sel4}
Gerwin Klein, June Andronick, Kevin Elphinstone, Toby Murray, Thomas Sewell,
  Rafal Kolanski, and Gernot Heiser.
\newblock Comprehensive formal verification of an {OS} microkernel.
\newblock {\em ACM Transactions on Computer Systems}, 32(1):2:1--2:70, February
  2014.

\bibitem{keystone}
Dayeol Lee, David Kohlbrenner, Shweta Shinde, Krste Asanovic, and Dawn Song.
\newblock Keystone: an open framework for architecting trusted execution
  environments.
\newblock In {\em EuroSys}. {ACM}, 2020.

\bibitem{crypto-cap}
Michael LeMay, Joydeep Rakshit, Sergej Deutsch, David~M. Durham, Santosh Ghosh,
  Anant Nori, Jayesh Gaur, Andrew Weiler, Salmin Sultana, Karanvir Grewal, and
  Sreenivas Subramoney.
\newblock Cryptographic capability computing.
\newblock In {\em MICRO-54: 54th Annual IEEE/ACM International Symposium on
  Microarchitecture}, MICRO '21, page 253–267, New York, NY, USA, 2021.
  Association for Computing Machinery.

\bibitem{pac-it-up}
Hans Liljestrand, Thomas Nyman, Kui Wang, Carlos~Chinea Perez, Jan-Erik Ekberg,
  and N.~Asokan.
\newblock {PAC} it up: Towards pointer integrity using {ARM} pointer
  authentication.
\newblock In {\em 28th USENIX Security Symposium (USENIX Security 19)}, pages
  177--194, Santa Clara, CA, August 2019. USENIX Association.

\bibitem{haskell2010}
Simon Marlow.
\newblock Haskell 2010 language report.
\newblock \url{https://www.haskell.org/onlinereport/haskell2010/}, 2010.

\bibitem{linear-unique}
Daniel Marshall, Michael Vollmer, and Dominic Orchard.
\newblock Linearity and uniqueness: An entente cordiale.
\newblock In {\em Programming Languages and Systems: 31st European Symposium on
  Programming, ESOP 2022, Held as Part of the European Joint Conferences on
  Theory and Practice of Software, ETAPS 2022, Munich, Germany, April 2–7,
  2022, Proceedings}, page 346–375, Berlin, Heidelberg, 2022.
  Springer-Verlag.

\bibitem{rust}
Nicholas~D. Matsakis and Felix S.~Klock II.
\newblock The rust language.
\newblock In Michael Feldman and S.~Tucker Taft, editors, {\em Proceedings of
  the 2014 {ACM} SIGAda annual conference on High integrity language
  technology, {HILT} 2014, Portland, Oregon, USA, October 18-21, 2014}, pages
  103--104. {ACM}, 2014.

\bibitem{cheri-provenance}
Alfredo Mazzinghi, Ripduman Sohan, and Robert N.~M. Watson.
\newblock Pointer provenance in a capability architecture.
\newblock In {\em Proceedings of the 10th USENIX Conference on Theory and
  Practice of Provenance}, TaPP'18, page~2, USA, 2018. USENIX Association.

\bibitem{sgx}
Frank McKeen, Ilya Alexandrovich, Alex Berenzon, Carlos~V. Rozas, Hisham Shafi,
  Vedvyas Shanbhogue, and Uday~R. Savagaonkar.
\newblock Innovative instructions and software model for isolated execution.
\newblock In {\em HASP@ISCA}, page~10. {ACM}, 2013.

\bibitem{e-lang}
Mark~S. Miller, E.~Dean Tribble, and Jonathan Shapiro.
\newblock Concurrency among strangers.
\newblock In Rocco De~Nicola and Davide Sangiorgi, editors, {\em Trustworthy
  Global Computing}, pages 195--229, Berlin, Heidelberg, 2005. Springer Berlin
  Heidelberg.

\bibitem{cap-myths}
Mark~S Miller, Ka-Ping Yee, Jonathan Shapiro, et~al.
\newblock Capability myths demolished.
\newblock Technical report, Technical Report SRL2003-02, Johns Hopkins
  University Systems Research~…, 2003.

\bibitem{mpk}
Intel® 64 and ia-32 architectures software developer manual, 2018.

\bibitem{arm-mpu}
Armv8-m memory model and memory protection user guide.
\newblock \url{https://developer.arm.com/documentation/107565/latest}, 2022.
\newblock Accessed on 27 Feb 2023.

\bibitem{intelmpx}
Intel® memory protection extensions enabling guide.
\newblock
  \url{https://www.intel.com/content/www/us/en/developer/articles/guide/intel-memory-protection-extensions-enabling-guide.html},
  2016.

\bibitem{arm-mte}
Armv8.5-a memory tagging extension white paper.
\newblock \url{https://developer.arm.com/documentation/102925/0100}, 2022.
\newblock Accessed on 30 January, 2023.

\bibitem{borrowing-permissions}
Karl Naden, Robert Bocchino, Jonathan Aldrich, and Kevin Bierhoff.
\newblock A type system for borrowing permissions.
\newblock In {\em Proceedings of the 39th Annual ACM SIGPLAN-SIGACT Symposium
  on Principles of Programming Languages}, POPL '12, page 557–570, New York,
  NY, USA, 2012. Association for Computing Machinery.

\bibitem{pac}
Learn the architecture: Providing protection for complex software.
\newblock \url{https://developer.arm.com/documentation/102433/latest/}, 2022.

\bibitem{nested-enclave}
Joongun Park, Naegyeong Kang, Taehoon Kim, Youngjin Kwon, and Jaehyuk Huh.
\newblock Nested enclave: Supporting fine-grained hierarchical isolation with
  sgx.
\newblock In {\em 2020 ACM/IEEE 47th Annual International Symposium on Computer
  Architecture (ISCA)}, pages 776--789, 2020.

\bibitem{game-of-threads}
Jose~Rodrigo Sanchez~Vicarte, Benjamin Schreiber, Riccardo Paccagnella, and
  Christopher~W. Fletcher.
\newblock {\em Game of Threads: Enabling Asynchronous Poisoning Attacks}, page
  35–52.
\newblock Association for Computing Machinery, New York, NY, USA, 2020.

\bibitem{pie}
Moritz Schneider, Aritra Dhar, Ivan Puddu, Kari Kostiainen, and Srdjan Čapkun.
\newblock Composite enclaves: Towards disaggregated trusted execution.
\newblock {\em IACR Transactions on Cryptographic Hardware and Embedded
  Systems}, 2022(1):630–656, Nov. 2021.

\bibitem{stktokens}
Lau Skorstengaard, Dominique Devriese, and Lars Birkedal.
\newblock Stktokens: Enforcing well-bracketed control flow and stack
  encapsulation using linear capabilities.
\newblock {\em Proc. ACM Program. Lang.}, 3(POPL), jan 2019.

\bibitem{clean-lang-uniqueness}
Sjaak Smetsers, Erik Barendsen, Marko C. J. D.~van Eekelen, and Marinus~J.
  Plasmeijer.
\newblock Guaranteeing safe destructive updates through a type system with
  uniqueness information for graphs.
\newblock In {\em Proceedings of the International Workshop on Graph
  Transformations in Computer Science}, page 358–379, Berlin, Heidelberg,
  1993. Springer-Verlag.

\bibitem{linear-cap-separation-logic}
Thomas Van~Strydonck, Frank Piessens, and Dominique Devriese.
\newblock Linear capabilities for fully abstract compilation of
  separation-logic-verified code.
\newblock {\em Proc. ACM Program. Lang.}, 3(ICFP), jul 2019.

\bibitem{linear-type}
Philip Wadler.
\newblock Linear types can change the world!
\newblock In {\em PROGRAMMING CONCEPTS AND METHODS}. North, 1990.

\bibitem{linear-logic-use}
Philip Wadler.
\newblock Is there a use for linear logic?
\newblock In {\em Proceedings of the 1991 ACM SIGPLAN Symposium on Partial
  Evaluation and Semantics-Based Program Manipulation}, PEPM '91, page
  255–273, New York, NY, USA, 1991. Association for Computing Machinery.

\bibitem{riscvpriv}
Andrew Waterman, Krste Asanovi\'c, and John Hauser.
\newblock {\em The RISC-V Instruction Set Manual: Volume II: Privileged
  Architecture}, 2021.

\bibitem{capsicum}
Robert N.~M. Watson, Jonathan Anderson, Ben Laurie, and Kris Kennaway.
\newblock Capsicum: Practical capabilities for {UNIX}.
\newblock In {\em 19th {USENIX} Security Symposium, Washington, DC, USA, August
  11-13, 2010, Proceedings}, pages 29--46. {USENIX} Association, 2010.

\bibitem{cheri}
Robert N.~M. Watson, Jonathan Woodruff, Peter~G. Neumann, Simon~W. Moore,
  Jonathan Anderson, David Chisnall, Nirav~H. Dave, Brooks Davis, Khilan Gudka,
  Ben Laurie, Steven~J. Murdoch, Robert~M. Norton, Michael Roe, Stacey~D. Son,
  and Munraj Vadera.
\newblock {CHERI:} {A} hybrid capability-system architecture for scalable
  software compartmentalization.
\newblock In {\em 2015 {IEEE} Symposium on Security and Privacy, {SP} 2015, San
  Jose, CA, USA, May 17-21, 2015}, pages 20--37. {IEEE} Computer Society, 2015.

\bibitem{asyncshock}
Nico Weichbrodt, Anil Kurmus, Peter Pietzuch, and R{\"u}diger Kapitza.
\newblock Asyncshock: Exploiting synchronisation bugs in intel sgx enclaves.
\newblock In Ioannis Askoxylakis, Sotiris Ioannidis, Sokratis Katsikas, and
  Catherine Meadows, editors, {\em Computer Security -- ESORICS 2016}, 2016.

\bibitem{cornucopia}
Nathaniel Wesley~Filardo, Brett~F. Gutstein, Jonathan Woodruff, Sam Ainsworth,
  Lucian Paul-Trifu, Brooks Davis, Hongyan Xia, Edward Tomasz~Napierala,
  Alexander Richardson, John Baldwin, David Chisnall, Jessica Clarke, Khilan
  Gudka, Alexandre Joannou, A.~Theodore~Markettos, Alfredo Mazzinghi, Robert~M.
  Norton, Michael Roe, Peter Sewell, Stacey Son, Timothy~M. Jones, Simon~W.
  Moore, Peter~G. Neumann, and Robert N.~M. Watson.
\newblock Cornucopia: Temporal safety for cheri heaps.
\newblock In {\em 2020 IEEE Symposium on Security and Privacy (SP)}, pages
  608--625, 2020.

\bibitem{mondrian}
Emmett Witchel, Josh Cates, and Krste Asanovi{\'c}.
\newblock Mondrian memory protection.
\newblock In {\em Proceedings of the 10th international conference on
  Architectural support for programming languages and operating systems}, pages
  304--316, 2002.

\bibitem{cheri-cap-model}
Jonathan Woodruff, Robert N.~M. Watson, David Chisnall, Simon~W. Moore,
  Jonathan Anderson, Brooks Davis, Ben Laurie, Peter~G. Neumann, Robert~M.
  Norton, and Michael Roe.
\newblock The {CHERI} capability model: Revisiting {RISC} in an age of risk.
\newblock In {\em {ISCA}}, pages 457--468. {IEEE} Computer Society, 2014.

\bibitem{cherivoke}
Hongyan Xia, Jonathan Woodruff, Sam Ainsworth, Nathaniel~Wesley Filardo,
  Michael Roe, Alexander Richardson, Peter Rugg, Peter~G. Neumann, Simon~W.
  Moore, Robert N.~M. Watson, and Timothy~M. Jones.
\newblock Cherivoke: Characterising pointer revocation using {CHERI}
  capabilities for temporal memory safety.
\newblock In {\em Proceedings of the 52nd Annual {IEEE/ACM} International
  Symposium on Microarchitecture, {MICRO} 2019, Columbus, OH, USA, October
  12-16, 2019}, pages 545--557. {ACM}, 2019.

\bibitem{elasticlave}
Jason~Zhijingcheng Yu, Shweta Shinde, Trevor Carlson, and Prateek Saxena.
\newblock Elasticlave: An efficient memory model for enclaves.
\newblock In {\em 31st USENIX Security Symposium (USENIX Security 22)}, Boston,
  MA, August 2022. USENIX Association.

\end{thebibliography}

\newpage

\appendix
\onecolumn
    \onecolumn
\section{Complete \codename{} Operational Semantics}
\label{appx:complete-model}

Table~\ref{table:op-semantics} describes the operational semantics of
the \codename{} model.
The definitions of auxiliary functions are listed in Table~\ref{table:aux-func}.
For simplicity, we omit the thread executing the instruction (i.e., $k$) from both
the instruction and the auxiliary function arguments.

{
    \centering
    \begin{longtable}{lp{.8\linewidth}} 
    \caption{Operational semantics of
        \codename{}.}\label{table:op-semantics}\endfirsthead
    \toprule
    \textbf{Instruction} & \textbf{State transition} \\ 
        & \textbf{Conditions} \\
    \midrule
    $\imov\ \rr_d\ \rr_s$ &
    $\mstate \mapsto \fUpdatePC(\mstate[\Theta.k.\theta.\rr_d \mapsto
        w][\Theta.k.\theta.\rr_s \mapsto \fMoved(w)])$ \\ & 
    $\mstate = \mstatefd \land \Theta.k.\theta.\rr_s = w$\\ 
    \hline
    $\ild\ \rr_d\ \rr_s$ &
    $\mstate \mapsto \fUpdatePC(\mstate[\Theta.k.\theta.\rr_d \mapsto
        w][mem.addr \mapsto \fMoved(w)])$ \\  &
    $\mstate = \mstatefd \land \Theta.k.\theta.\rr_s = c \land \fValidCap(rt, c) \land
    \neg\fRevoked(rt, c) \land \fInBoundCap(c) \land \fAccessibleCap(c) \land
    \fReadableCap(c) \land c.a = addr \land mem.addr = w \land
        (\fLinearCap(w) \implies \fWritableCap(c))$ \\
    \hline
    $\isd\ \rr_d\ \rr_s$ &
    $\mstate \mapsto \fUpdatePC(\mstate[mem.addr \mapsto
        w][\Theta.k.\theta.\rr_s \mapsto \fMoved(w)][\Theta.k.\theta.\rr_d \mapsto
        \fUpdateCursor(c)])$ \\ & 
    $\mstate = \mstatefd \land \Theta.k.\theta.\rr_d = c \land \fValidCap(rt, c) 
    \land \fInBoundCap(c) \land \fAccessibleCap(c) \land
    \fWritableCap(c) \land c.a = addr \land \Theta.k.\theta.\rr_s = w$ \\
    \hline
    $\itighten\ \rr_d\ \rr_s$ &
    $\mstate \mapsto \fUpdatePC(\mstate[\Theta.k.\theta.\rr_d \mapsto
    c^\prime])$ \\ &
    $\mstate = \mstatefd \land \Theta.k.\theta.\rr_d = c \land \fValidCap(rt, c) \land c = \capfd \land
    \Theta.k.\theta.\rr_s = n \in \fullN \land c^\prime = (t, b, e, a, \fTightenPerm(p, \fDecodePerm(n)))$\\
    \hline
    $\ishrink\ \rr_d\ \rr_b\ \rr_e$ &
    $\mstate \mapsto \fUpdatePC(\mstate[\Theta.k.\theta.\rr_d \mapsto
    c^\prime])$ \\ &
    $\mstate = \mstatefd \land \Theta.k.\theta.\rr_b = b^\prime \in \fullN \land \Theta.k.\theta.\rr_e = e^\prime \in \fullN \land \Theta.k.\theta.\rr_d = c \land \fValidCap(rt, c) \land c = \capfd \land
    t \in \{\tLin, \tNon\} \land b \leq b^\prime < e^\prime \leq e \land c^\prime = (t, b^\prime, e^\prime, a, p, n)$
    \\
    \hline
    $\isplit\ \rr_d\ \rr_s\ \rr_p$ &
    $\mstate \mapsto \fUpdatePC(\mstate[\Theta.k.\theta.\rr_d \mapsto
    c^\prime_1][\Theta.k.\theta.\rr_s \mapsto c^\prime_2][rt \mapsto
    rt^\prime])$ \\ &
    $\mstate = \mstatefd \land \Theta.k.\theta.\rr_d = c \land \fValidCap(rt, c)  \land
    c = (\tLin, b, e, a, p, n) \land \Theta.k.\theta.\rr_p = pv \land b < pv < e \land c^\prime_1 = (\tLin, b, pv, a, p, n)
    \land c^\prime_2 = (\tLin, pv, e, a, p, n^\prime) \land n^\prime \in \fullN \land n^\prime \not\in \mathrm{dom}(rt) \land rt^\prime = rt[n^\prime \mapsto rt.n]$ 
    \\
    \hline
    $\idelin\ \rr$ & 
    $\mstate \mapsto \fUpdatePC(\mstate[\Theta.k.\theta.\rr \mapsto c^\prime,
    rt \mapsto rt^\prime])$ \\ &
    $\mstate = \mstatefd \land \Theta.k.\theta.\rr = c \land \fValidCap(rt, c) \land c = (\tLin, b, e, a, p, n) \land
    c^\prime = (\tNon, b, e, a, p, n) \land rt.n = (n^\prime, nt) \land rt^\prime = rt[n
        \mapsto (n^\prime, \tRNon)]$ \\
    \hline
    $\iscc\ \rr_d\ \rr_s$ & 
    $\mstate \mapsto \fUpdatePC(\mstate[\Theta.k.\theta.\rr_d \mapsto c^\prime])$ \\
    &
    $\mstate = \mstatefd \land \Theta.k.\theta.\rr_s = n \land n \in \fullN
    \land \Theta.k.\theta.\rr_d = c \land c \in \sCap \land
    c^\prime = c[a \mapsto n]$
    \\ \hline
    $\ilcc\ \rr_d\ \rr_s$ &
    $\mstate \mapsto \fUpdatePC(\mstate[\Theta.k.\theta.\rr_d \mapsto n])$ \\ &
    $\mstate = \mstatefd \land \Theta.k.\theta.\rr_s = c \land c \in \sCap \land
    n = c.a$ \\
    \hline
    $\irevoke\ \rr$ & 
    $\mstate \mapsto \fUpdatePC(\mstate[\Theta.k.\theta.\rr \mapsto
    c^\prime][rt \mapsto rt^\prime])$ \\ &
    $\mstate = \mstatefd \land \Theta.k.\theta.\rr = c \land \fValidCap(rt, c) \land c = (\tRev, b, e, a, p, n) \land rt^\prime = \fReparent(rt, n, \nnull)
    \land c^\prime = \begin{cases}
    (\tLin,
        b, e, a, p, n) & (\fullN \times \{(n, \tRLin)\} \cap rt) =
        \emptyset \lor p \in \{\pNA, \pR, \pRX\}\\
    (\tUninit, b, e, b, p, n) & \text{otherwise}
    \end{cases}$
    \\
    \hline
    $\imrev\ \rr_d\ \rr_s$ &
    $\mstate \mapsto \fUpdatePC(\mstate[rt \mapsto
    rt^\prime][\Theta.k.\theta.\rr_d \mapsto c^\prime])$ \\ &
    $\mstate = \mstatefd \land \Theta.k.\theta.\rr_s = c \land \fValidCap(rt, c) \land c = (\tLin, b, e, a, p, n) \land
    n^\prime \in \fullN \land n^\prime \not\in \mathrm{dom}(rt) \land
    c^\prime = (\tRev, b, e, a, p, n^\prime) \land
    rt^\prime = (rt[n \mapsto (n^\prime, \tRLin)]) \cup \{(n^\prime, rt.n)\}$
    \\
    \hline
    $\iinit\ \rr$ &
    $\mstate \mapsto \fUpdatePC(\mstate[\Theta.k.\theta.\rr = c^\prime])$ \\ &
    $\mstate = \mstatefd \land \Theta.k.\theta.\rr = c \land \fValidCap(rt, c) \land c = (\tUninit, b, e, a, p, n) \land a = e \land c^\prime = (\tLin, b, e, a, p, n)$
    \\
    \hline
    $\idrop\ \rr$ &
    $\mstate \mapsto \fUpdatePC(\mstate[rt \mapsto
    rt^\prime][\Theta.k.\theta.\rr \mapsto 0])$ \\ &
    $\mstate = \mstatefd \land \Theta.k.\theta.\rr = c \land \fValidCap(rt, c) \land c = \capfd \land
    t \in \{\tLin, \tRev, \tUninit, \tSealed, \tSealedRet\} \land rt^\prime = \fRemove(\fReparent(rt, n, rt.n), n)$
    \\
    \hline
    $\iseal\ \rr$ &
    $\mstate \mapsto \fUpdatePC(\mstate[\Theta.k.\theta.\rr \mapsto
        c^\prime][N \mapsto \mstate.N + 1]])$ \\ & 
    $\mstate = \mstatefd \land \Theta.k.\theta.\rr = c \land \fValidCap(rt, c) \land
    \fReadableCap(c) \land \fWritableCap(c) \land
    c = (\tLin, b, e, a, p, n) \land c^\prime = (\tSealed(\mstate.N), b, e, a, p, n)
    $
    \\
    \hline
    $\icall\ \rr_d\ \rr_s$ &
    $\mstate \mapsto \mstate[\Theta.k \mapsto (regs^\prime, d)][mem \mapsto
        mem^\prime]$ \\ & 
    $\mstate = \mstatefd \land \Theta.k.\theta.\rr_d = c_i \land 
    \Theta.k.\theta.\rr_s = w \land 
    \fValidCap(rt, c_i) \land c_i = (\tSealed(d), b_i, e_i, a_i, p_i, n_i) \land
    b_i + M + 4 \leq e_i \land
    regs^\prime = \fLoadContext(\Theta.k.\theta, b_i, mem)
    [\rret \mapsto c_i[t \mapsto
    \tSealedRet(\mstate.k.d, \rr_d)]][\rr_1 \mapsto w]
    \land
    mem^\prime = \fSaveContext(mem, b_i, \Theta.k.\theta[\rr_s \mapsto
    \fMoved(w)][\rr_d \mapsto 0])$
    \\
    \hline
    $\ireturn\ \rr_d\ \rr_s$ &
    $\mstate \mapsto \mstate[\Theta.k \mapsto (regs^\prime, d)][mem \mapsto
        mem^\prime]$ \\ & 
    $\mstate = \mstatefd \land \Theta.k.\theta.\rr_d = c_i \land
    \Theta.k.\theta.\rr_s = w \land 
    \fValidCap(rt, c_i) \land c_i = (\tSealedRet(d, \rr), b_i, e_i, a_i, p_i, n_i) \land
    b_i + M + 4 \leq e_i \land
    regs^\prime = \fLoadContext(regs, mem, b_i)
    [\rr \mapsto w] \land
    mem^\prime = \fClearSealed(mem, b_i)$
    \\ \hline
    $\iretseal\ \rr_d\ \rr_s$ &
    $\mstate \mapsto \mstate[\Theta.k \mapsto (regs^\prime, d)][mem \mapsto
    mem^\prime]$ \\ &
    $\mstate = \mstatefd \land \Theta.k.\theta.\rr_d = c_i \land
    \Theta.k.\theta.\rr_s = w \land 
    \fValidCap(rt, c_i) \land c_i = (\tSealedRet(d, \rr), b_i, e_i, a_i, p_i, n_i) \land
    b_i + M + 4 \leq e_i \land
    regs^\prime = \fLoadContext(\Theta.k.\theta, mem, b_i)
    [\rr \mapsto c_i[t \mapsto \tSealed(\mstate.k.d)]] \land
    mem^\prime = \fSaveContext(mem, b_i, \Theta.k.\theta[\rpc.a
    \mapsto w][\rr_d \mapsto 0])$
    \\ \hline
    $\iexcept\ \rr$ & $= \icall\ \repc\ \rr\ \text{(without incrementing PC)}$ \\ \hline
    $\ijmp\ \rr$ &
    $\mstate = \mstate[\Theta.k.\theta.\rpc \mapsto c^\prime]$ \\ &
    $\mstate = \mstatefd \land \Theta.k.\theta.\rr = n \land n \in \fullN \land
    c^\prime = \Theta.k.\theta.\rpc[a \mapsto n]$
    \\ \hline

    $\ijnz\ \rr_d\ \rr_s$ &
    $\mstate = \mstate[\Theta.k.\theta.\rpc \mapsto c^\prime]$ \\ &
    $\mstate = \mstatefd \land \Theta.k.\theta.\rr_s = n_s \land \Theta.k.\theta.\rr_d = n_d \land n_s, n_d \in \fullN \land c^\prime = \begin{cases}
    \fIncrementCursor(\Theta.k.\theta.\rpc) & n_s = 0\\
    \Theta.k.\theta.\rpc[a \mapsto n_d] & \text{otherwise}
    \end{cases}$
    \\ \hline
    $\ili\ \rr\ n$ &
    $\mstate \mapsto \fUpdatePC(\mstate[\Theta.k.\theta.\rr \mapsto n])$ \\ \hline
    $\iadd\ \rr_d\ \rr_s$ &
    $\mstate \mapsto \fUpdatePC(\mstate[\Theta.k.\theta.\rr_d \mapsto
    n_d^\prime])$ \\ &
    $\mstate = \mstatefd \land \Theta.k.\theta.\rr_d = n_d \land \Theta.k.\theta.\rr_s = n_s \land
    n_d, n_s \in \fullN \land n_d^\prime = n_d + n_s$
    \\ \hline
    $\ilt\ \rr_d\ \rr_a\ \rr_b$ &
    $\mstate = \fUpdatePC(\mstate[\Theta.k.\theta.\rr_d \mapsto n])$ \\ &
    $\mstate = \mstatefd \land \Theta.k.\theta.\rr_a = n_a \land \Theta.k.\theta.\rr_b = n_b \land n_a, n_b \in \fullN \land n = \begin{cases}
    1 & n_a < n_b\\
    0 & \text{otherwise}
    \end{cases}$
    \\ \hline
    $\iinvalid$ &
    $\mstate \mapsto \mstate[\Theta.k.\theta \mapsto \aerror]$ \\
    \bottomrule
    \end{longtable}
    }

\begin{table*}[t]
\centering
\caption{Auxiliary functions used in the operational semantics definition of \codename{}. $k$ is the currently executing thread, and is omitted in the function arguments below for simplicity. It will be supplied as the first
argument when necessary.}
\label{table:aux-func}
    \begin{tabularx}{\linewidth}{lX}
    \hline
    \textbf{Function} & \textbf{Definition} \\
    \hline
        $\fLinearTypes$ & $\{\tLin, \tUninit, \tRev, \tSealed,
        \tSealedRet\}$ \\ \hline
       
    $\fUpdatePC(\mstate)$ & $\begin{cases} \mstate[\Theta.k.\theta.\rpc.a \mapsto a + 1] & \mstate = \mstatefd \land \Theta.k.\theta.\rpc = \capfd \\ \aerror & \text{otherwise} \end{cases}$ \\
    \hline
    $\fValidPC(\mstate)$ & $\mstate = \mstatefd \land \Theta.k.\theta.\rpc \in \sCap \land \fExecutableCap(\Theta.k.\theta.\rpc) \land \fValidCap(rt, \Theta.k.\theta.\rpc) \land \fInBoundCap(\Theta.k.\theta.\rpc) \land \fAccessibleCap(\Theta.k.\theta.\rpc)$ \\
    \hline
    $\fReadableCap(c)$ & $c = \capfd \land p \in \{\pR, \pRX, \pRWX\} \land t \neq \tUninit$ \\
    \hline
    $\fWritableCap(c)$ & $c = \capfd \land p \in \{\pRW, \pRWX\}$ \\
    \hline
    $\fExecutableCap(c)$ & $c = \capfd \land p \in \{\pRX, \pRWX\} \land t \neq \tUninit$\\
    \hline
    $\fValidCap(rt, c)$ & $c = \capfd \land \neg\fRevoked(rt, n)$ \\
    \hline
    $\fRevoked(rt, n)$ & $
    \begin{cases}
    \cfalse & rt.n.pr = \nroot \\
    \ctrue & rt.n.pr = \nnull \\
    \fRevoked(rt, rt.n) & \text{otherwise}
    \end{cases}
    $ \\
    \hline
    $\fInBoundCap(c)$ & $c = \capfd \land b \leq a < e$ \\
    \hline
    $\fAccessibleCap(c)$ & $c = \capfd \land t \in \{\tLin, \tNon, \tUninit\}$ \\
    \hline
    $\fLinearCap(c)$ & $c = \capfd \land t \in \fLinearTypes$ \\
    \hline
   
    $\fMoved(w)$ & $
    \begin{cases}
        0 & \fLinearCap(w) \\
    w & \text{otherwise}
    \end{cases}
    $ \\
    \hline
    $\fReparent(rt, n, n^\prime)$ &
    $(rt - \fullN \times \{n\} \times \sRNodeType) \cup \{(k,
        n^\prime, nt) \mid (k, n, nt) \in rt\}$ \\
    \hline
    $\fRemove(rt, n)$ &
        $rt - \{n\} \times \sRevParent \times \sRNodeType$ \\ \hline
    $\fDecodePerm(n)$ &
    $\begin{cases}
    \pR & 0 \\
    \pRW & 1 \\
    \pRX & 2 \\
    \pRWX & 3 \\
    \pNA & \text{otherwise}
    \end{cases}$ \\ \hline
    $\fTightenPerm(p, p^\prime)$  &
    $
    \begin{cases}
    p^\prime & p^\prime \preceq p \\
    \pNA & \text{otherwise}
    \end{cases}
    $
    \\ \hline
    $\fIncrementCursor(c)$ &
    $c[a \mapsto c.a + 1]$ \\ \hline
    $\fUpdateCursor(c)$ &
    $\begin{cases}
    c & c \in \sCap\ \land c.t \neq \tUninit \\
    \fIncrementCursor(c) & c \in \sCap\ \land c.t = \tUninit \\
    0 & \text{otherwise}
    \end{cases}$ \\ \hline
    $\fFetchInsn(\mstate)$ & 
    $\begin{cases}
    i & \Theta.k.\theta \in\sRegFile \land \fValidPC(\mstate) \land \mstate.\Theta.k.\theta.\rpc = \capfd \land i = \mstate.mem.a \land i \in \sInsn \\
    \iinvalid & \text{otherwise}
    \end{cases}$ 
    \\  \hline
    $\fLoadContext(regs, mem, b)$ & $regs[\rpc \mapsto mem.b][\repc \mapsto mem.(b + 2)][\rret \mapsto mem.(b +
    3)][\rr_1 \mapsto
    mem.(b + 4)][\rr_2 \mapsto mem.(b + 5)]\cdots[\rr_M \mapsto mem.(b + M + 3)]$ \\
    \hline
    $\fSaveContext(mem, b, regs)$ & $mem[b \mapsto regs.\rpc][b + 2 \mapsto
    regs.\repc][b + 3 \mapsto regs.\rret][b + 4 \mapsto regs.\rr_1][b
    + 5 \mapsto regs.\rr_2]\cdots[b + M + 3 \mapsto regs.\rr_M]$ \\
    \hline
    $\fClearSealed(mem, b)$ & $mem[b \mapsto 0][b + 1 \mapsto 0]
    \cdots[b + M + 3 \mapsto 0]$ \\ \hline
    \end{tabularx}
\end{table*}

The relation $\preceq \subseteq \sPerms \times \sPerms$ is defined as
\begin{equation}
\begin{array}{cl}
    \preceq = &
     \{\pNA\} \times \sPerms\ \cup \\
     & \{\pR\} \times \{\pR, \pRW, \pRX, \pRWX\} \cup \\
     & \{\pRW\} \times \{\pRW, \pRWX\} \cup \\
     & \{\pRX\} \times \{\pRX, \pRWX\} \cup \{(\pRWX, \pRWX)\}
\end{array}
\end{equation}

\onecolumn
\section{\abscodename{}}
\label{sec:absproofs}

\subsection{Notations and Definitions}

For any set $S \subset \fullN$, we use $\overline{S}$
to denote its complement, i.e., $\fullN - S$.

We define the state transition function below.


\begin{definition}[$F, R$]
    $F: \sState \times \fullN \to \sState \times \fullN$:
    \begin{equation}
        F(\mstate, \sigma) = (\mstate^\prime, \sigma^\prime),
    \end{equation}
    where
    \begin{equation}
        \mstate^\prime = \begin{cases}
            \fExecute(\mstate, k, \fFetchInsn(\mstate, k)) & s = 0 \\
            \fExecute(\mstate, k, \iexcept) & \text{otherwise,}
        \end{cases}
    \end{equation}
    and
    \begin{equation}
        (s, k, \sigma^\prime) = R(\sigma),
    \end{equation}
    where
    \begin{equation}
        R: \fullN \to \{0, 1\} \times \fullN \times \fullN.
    \end{equation}
\end{definition}

\begin{definition}[Auxiliary functions]
    We define the following functions:
\begin{itemize}
    \item $\fSealedRegs(c) = \{(\rpc, c.b), (\repc, c.b + 2), (\rret,
        c.b + 3), (\rr_1, c.b + 4), \cdots, (\rr_M, c.b + M + 3)\}$.
\end{itemize}
\end{definition}

\begin{definition}[$\sDLoc$]
We use $\sDLoc$ to denote the location of a piece of data.
There are two possibilities: it can be a memory location identified by
its address or a register identified by the corresponding domain and
    the register name, hence
\begin{equation}
    \sDLoc \ni loc \coloneqq \sDLocMem(n) \mid \sDLocReg(n, r).
\end{equation}

\end{definition}

\subsection{Abstract Model}
\vspace{1\baselineskip}
\begin{minipage}{0.5\textwidth}
\begin{itemize}
    \item $\textit{dom} \coloneqq \textsf{user $\mid$ sup $\mid$ sub}$
\end{itemize}

\begin{itemize}
    \item $\textit{act}_\textit{abs} \coloneqq \\
    \textsf{load\_linear } \textit{addr } \textit{payload} \mid \\
    \textsf{store\_linear } \textit{addr } \textit{payload} \mid \\
    \textsf{load } \textit{addr } \textit{payload} \mid \\
    \textsf{store } \textit{addr } \textit{payload} \mid \\
    \textsf{split } \textit{cap nat} \mid \\
    \textsf{shrink } \textit{cap nat} \mid \\
    \textsf{send } \textit{cap}~\textit{dom} \mid \\
    \textsf{discard } \textit{cap} \mid \\
    \textsf{claim } \textit{cap} \mid \\
    \textsf{revoke } \textit{cap}$ \\
\end{itemize}
\end{minipage}
\begin{minipage}{0.5\textwidth}
\begin{itemize}
    \item $\textit{mem}_{\textit{abs}} \coloneqq \sAddr \mapsto (\sWord \mid \textsf{uninit})$
\end{itemize}

\begin{itemize}
    \item $\textit{range} \coloneqq \{n \mid x \leq n < x+y \}$
\end{itemize}

\begin{itemize}
    \item $\textit{payload} \coloneqq \text{Word}$
    \item $\textit{cap}_{\textit{abs}} \coloneqq \textit{range}$
    \item $\textit{tstate} \coloneqq \textit{cap$_\textit{abs}$ set}$
    \item $\textit{pstate} \coloneqq (\textit{mem}_\textit{abs}, \textit{tstate}_\textsf{user}, \textit{tstate}_\textsf{sup},  \textit{tstate}_\textsf{sub})$\\
\end{itemize}
\end{minipage}

\subsection{Abstract Semantics}
\label{appx:abs_semantics}

$\text{Owns}(\textit{tstate}, \textit{a}) = \exists \textit{c}.~\textit{c} \in \textit{tstate} \wedge \textit{a} \in \textit{c}$\\

$\text{Split}(~\{n \mid x \leq n < x+y \}~,~\textit{a}) = \{~\{n \mid x \leq n < a \},~\{n \mid a \leq n < x+y \}~\}$\\

$\textit{pstate} \hookrightarrow^{\textit{act}_\textit{abs}}_\textit{dom} \textit{pstate}'$\\

\noindent load linear:\\
$(\textit{mem}_\textit{abs}, \textit{tstate}_\textsf{user}, \textit{tstate}_\textsf{sup},  \textit{tstate}_\textsf{sub}) \hookrightarrow^{\textsf{load\_linear } \textit{addr payload}}_\textit{dom} (\textit{mem}_\textit{abs}, \textit{tstate}_\textsf{user}, \textit{tstate}_\textsf{sup},  \textit{tstate}_\textsf{sub})$\\[0.2\baselineskip]
\indent\indent if $\textit{payload} = \textit{mem}_\textit{abs}[\textit{addr}]~\wedge$\\ \indent\indent\indent $\text{Owns}(\textit{tstate}_\textit{dom}, \textit{addr})$\\

\noindent store linear:\\
$(\textit{mem}_\textit{abs}, \textit{tstate}_\textsf{user}, \textit{tstate}_\textsf{sup},  \textit{tstate}_\textsf{sub}) \hookrightarrow^{\textsf{store\_linear } \textit{addr payload}}_\textit{dom} (\textit{mem}_\textit{abs}', \textit{tstate}_\textsf{user}, \textit{tstate}_\textsf{sup},  \textit{tstate}_\textsf{sub})$\\[0.2\baselineskip]
\indent\indent if $\textit{mem}_\textit{abs}' = \textit{mem}_\textit{abs}[\textit{addr} := \textit{payload}]~\wedge$\\ \indent\indent\indent $\text{Owns}(\textit{tstate}_\textit{dom}, \textit{addr})$\\

\noindent load:\\
$(\textit{mem}_\textit{abs}, \textit{tstate}_\textsf{user}, \textit{tstate}_\textsf{sup},  \textit{tstate}_\textsf{sub}) \hookrightarrow^{\textsf{load } \textit{addr payload}}_\textit{dom} (\textit{mem}_\textit{abs}, \textit{tstate}_\textsf{user}, \textit{tstate}_\textsf{sup},  \textit{tstate}_\textsf{sub})$\\[0.2\baselineskip]
\indent\indent if $\textit{payload} = \textit{mem}_\textit{abs}[\textit{addr}]~\wedge$\\
\indent\indent\indent $\nexists \textit{dom}'.~\text{Owns}(\textit{tstate}_{\textit{dom}'}, \textit{addr})$\\

\noindent store:\\
$(\textit{mem}_\textit{abs}, \textit{tstate}_\textsf{user}, \textit{tstate}_\textsf{sup},  \textit{tstate}_\textsf{sub}) \hookrightarrow^{\textsf{store\_linear } \textit{addr payload}}_\textit{dom} (\textit{mem}_\textit{abs}', \textit{tstate}_\textsf{user}, \textit{tstate}_\textsf{sup},  \textit{tstate}_\textsf{sub})$\\[0.2\baselineskip]
\indent\indent if $\textit{mem}_\textit{abs}' = \textit{mem}_\textit{abs}[\textit{addr} := \textit{payload}]~\wedge$\\
\indent\indent\indent $\nexists \textit{dom}'.~\text{Owns}(\textit{tstate}_{\textit{dom}'}, \textit{addr})$\\

\noindent split:\\
$(\textit{mem}_\textit{abs}, \textit{tstate}_\textsf{user}, \textit{tstate}_\textsf{sup},  \textit{tstate}_\textsf{sub}) \hookrightarrow^{\textsf{split } \textit{cap n}}_\textit{dom} (\textit{mem}_\textit{abs}, \textit{tstate}_\textsf{user}', \textit{tstate}_\textsf{sup}',  \textit{tstate}_\textsf{sub}')$\\[0.2\baselineskip]
\indent\indent if $\textit{cap} \in \textit{tstate}_{\textit{dom}}~\wedge$\\
\indent\indent\indent $\textit{nat} \in \textit{cap}~\wedge$\\
\indent\indent\indent $\textit{tstate}_{\textit{dom}}' = (\textit{tstate}_{\textit{dom}} - \{\textit{cap}\}) \cup \text{Split}(\textit{cap},\textit{n})~\wedge$\\
\indent\indent\indent $\forall \textit{dom}'\neq \textit{dom}.~\textit{tstate}_\textit{dom$'$}' = \textit{tstate}_\textit{dom$'$}$\\

\noindent shrink:\\
$(\textit{mem}_\textit{abs}, \textit{tstate}_\textsf{user}, \textit{tstate}_\textsf{sup},  \textit{tstate}_\textsf{sub}) \hookrightarrow^{\textsf{split } \textit{cap n}}_\textit{dom} (\textit{mem}_\textit{abs}, \textit{tstate}_\textsf{user}', \textit{tstate}_\textsf{sup}',  \textit{tstate}_\textsf{sub}')$\\[0.2\baselineskip]
\indent\indent if $\textit{cap} \in \textit{tstate}_{\textit{dom}}~\wedge$\\
\indent\indent\indent $\textit{nat} \in \textit{cap}~\wedge$\\
\indent\indent\indent $\textit{tstate}_{\textit{dom}}' = (\textit{tstate}_{\textit{dom}} - \{\textit{cap}\}) \cup \text{Shrink}(\textit{cap},\textit{n})~\wedge$\\
\indent\indent\indent $\forall \textit{dom}'\neq \textit{dom}.~\textit{tstate}_\textit{dom$'$}' = \textit{tstate}_\textit{dom$'$}$\\

\noindent send:\\
$(\textit{mem}_\textit{abs}, \textit{tstate}_\textsf{user}, \textit{tstate}_\textsf{sup},  \textit{tstate}_\textsf{sub}) \hookrightarrow^{\textsf{send } \textit{cap dom$'$}}_\textit{dom} (\textit{mem}_\textit{abs}, \textit{tstate}_\textsf{user}', \textit{tstate}_\textsf{sup}',  \textit{tstate}_\textsf{sub}')$\\[0.2\baselineskip]
\indent\indent if $\textit{cap} \in \textit{tstate}_{\textit{dom}}~\wedge$\\
\indent\indent\indent $\textit{tstate}_{\textit{dom}}' = \textit{tstate}_{\textit{dom}} - \{\textit{cap}\}~\wedge$\\ \indent\indent\indent $\textit{tstate}_{\textit{dom}'}' = \textit{tstate}_{\textit{dom}'} \cup \{\textit{cap}\}~\wedge$\\
\indent\indent\indent $\textit{tstate}_{\textit{dom}'' \neq \textit{dom} \neq \textit{dom}'}' = \textit{tstate}_{\textit{dom}''}$\\

\noindent discard:\\
$(\textit{mem}_\textit{abs}, \textit{tstate}_\textsf{user}, \textit{tstate}_\textsf{sup},  \textit{tstate}_\textsf{sub}) \hookrightarrow^{\textsf{discard } \textit{cap}}_\textit{dom} (\textit{mem}_\textit{abs}, \textit{tstate}_\textsf{user}', \textit{tstate}_\textsf{sup}',  \textit{tstate}_\textsf{sub}')$\\[0.2\baselineskip]
\indent\indent if $\textit{cap} \in \textit{tstate}_{\textit{dom}}~\wedge$\\
\indent\indent\indent $\textit{tstate}_{\textit{dom}}' = \textit{tstate}_{\textit{dom}} - \{\textit{cap}\}~\wedge$\\ \indent\indent\indent $\forall \textit{dom}'\neq \textit{dom}.~\textit{tstate}_\textit{dom$'$}' = \textit{tstate}_\textit{dom$'$}$\\

\noindent claim:\\
$(\textit{mem}_\textit{abs}, \textit{tstate}_\textsf{user}, \textit{tstate}_\textsf{sup},  \textit{tstate}_\textsf{sub}) \hookrightarrow^{\textsf{claim } \textit{cap}}_\textit{dom} (\textit{mem}_\textit{abs}, \textit{tstate}_\textsf{user}', \textit{tstate}_\textsf{sup}',  \textit{tstate}_\textsf{sub}')$\\[0.2\baselineskip]
\indent\indent if $\textit{cap} \notin \textit{tstate}_{\textsf{user}\cup\textsf{sub}\cup\textsf{sup}}~\wedge$\\
\indent\indent\indent $\textit{tstate}_{\textit{dom}}' = \textit{tstate}_{\textit{dom}} \cup \{\textit{cap}\}~\wedge$\\ \indent\indent\indent $\forall \textit{dom}'\neq \textit{dom}.~\textit{tstate}_\textit{dom$'$}' = \textit{tstate}_\textit{dom$'$}$\\

\noindent revoke:\\
$(\textit{mem}_\textit{abs}, \textit{tstate}_\textsf{user}, \textit{tstate}_\textsf{sup},  \textit{tstate}_\textsf{sub}) \hookrightarrow^{\textsf{revoke } \textit{cap}}_\textit{dom} (\textit{mem}_\textit{abs}', \textit{tstate}_\textsf{user}', \textit{tstate}_\textsf{sup}',  \textit{tstate}_\textsf{sub}')$\\[0.2\baselineskip]
\indent\indent if $
\textit{tstate}_\textit{dom}' = \textit{tstate}_\textit{dom} \cup \{ \textit{cap} \}~\wedge$\\
\indent\indent\indent $\forall \textit{dom}'\neq \textit{dom}.~\textit{tstate}_\textit{dom$'$}' = \textit{tstate}_\textit{dom$'$} - \{ \textit{cap} \}~\wedge$\\
\indent\indent\indent $(\textit{dom} = \textsf{sub} \longrightarrow \textit{cap} \notin \textit{tstate}_\textsf{user})~\wedge$\\
\indent\indent\indent$(\textit{mem}_\textit{abs}' = \textit{mem}_\textit{abs}[\textit{cap} := \textsf{uninit}])$\\

An intuitive summary of the semantics of each abstract is provided in
Table~\ref{table:abstract-actions}.

\begin{table}[t]
    \caption{Actions in \abscodename{}.}
    \label{table:abstract-actions}
    \begin{center}
    {\begin{tabular}{ll}
        \hline
        \textbf{Action} & \textbf{Description} \\
        \hline
    $\textsf{load\_linear } \textit{addr } \textit{payload}$  & load \textit{payload} from memory \\
    $\textsf{store\_linear } \textit{addr } \textit{payload}$ & store \textit{payload} to memory  \\
    $\textsf{load } \textit{addr } \textit{payload} $  & load \textit{payload} from memory \\
    $\textsf{store } \textit{addr } \textit{payload}$ & store \textit{payload} to memory \\
    $\textsf{split } \textit{cap nat}$ & split \textit{cap} in two at \textit{nat} \\
    $\textsf{shrink } \textit{cap nat}$ & shrink \textit{cap} to length \textit{nat} \\
    $\textsf{send } \textit{cap}~\textit{dom}$ & send \textit{cap} to another domain \\
    $\textsf{discard } \textit{cap} $ & remove \textit{cap} from the domain \\
    $\textsf{claim } \textit{cap}$  & take ownership of unowned
        \textit{cap}  \\[.05cm]
    $\textsf{revoke } \textit{cap}$ & \begin{tabular}{@{}l}wrest ownership of \textit{cap},\\ setting relevant memory to \textsf{uninit}\end{tabular} \\
    \hline
    \end{tabular}}
    \end{center}
\end{table}

\subsection{Refinement Mapping}

Function from $\mstate$ to \textit{pstate}, with respect to a distinguished user domain $\textit{d} \in \mathbb{N}$ and a distinguished set of subordinate domains $\textit{D}_\textsf{sub} \subseteq (\mathbb{N} - \textit{d})$.\\

\noindent $\text{Refines}_\textit{(d,$D_\textsf{sub}$)}(\mstate, (\textit{mem}_\textit{abs}, \textit{tstate}_\textsf{user}, \textit{tstate}_\textsf{sup},  \textit{tstate}_\textsf{sub}))$

\noindent where:\\
\indent  $\textit{ind}_\textit{uninit} = \bigcup_{d \in \mathbb{N}}~\fworanges(\realm_w (\mstate, d))$ \\
\indent $\textit{mem}_\textit{abs} = \mstate.\textit{mem}[\textit{ind}_\textit{uninit} := \textsf{uninit}]$\\
\indent $\textit{tstate}_\textsf{user} = \text{ranges}(\xrealm(\mstate, d))$\\
\indent$\textit{tstate}_\textsf{sub} = \bigcup_{d' \in D_\textsf{sub}}~\text{ranges}(\xrealm(\mstate, d'))$\\
\indent$\textit{tstate}_\textsf{sup} = \bigcup_{d' \in (\mathbb{N} - D_\textsf{sub} - d)}~\text{ranges}(\xrealm(\mstate, d'))$\\
$\text{Sub}(\mstate,D_\textsf{sub})$\\

\subsubsection{Subordinate Environment Invariant}

\noindent $\text{Sub}(\mstate,D_\textsf{sub})$

\noindent where:\\
\indent $\forall c \in \xrealm(\mstate, d).~\nexists c' \in (\bigcup_{k \in \textit{D}_\textsf{sub}} C(\mstate,k)).~c'.t = \tRev \wedge \text{range}(c) \cup \text{range}(c') \neq \emptyset$\\

Intuitively, a subordinate domain may never locally hold a revocation capability for a capability in the exclusive realm of the user domain.
This is an \textit{assumption} rather than a proof obligation.
By default, all domains other than the user domain must be assumed to be part of the superordinate environment, unless the user has some trustable information regarding their behaviour (for example, because it created the other domain itself).\\

Table~\ref{table:concrete-abstract} presents a summarizing sketch of
the mapping from instructions in the concrete model to abstract
actions.

\begin{table}[t]
    \centering
    \caption{Mapping from instructions to abstract actions.}
    \label{table:concrete-abstract}
    \begin{tabular}{l|l|l|l}
    \hline
        \textbf{Instruction} & $\irevoke\ \rr$ & $\idelin\ \rr$ &
        $\ishrink\ \rr_d\ \rr_b\ \rr_e$ \\
        \textbf{Action} & \textsf{revoke} &  \textsf{discard} & 
        \textsf{shrink}\\
    \hline
         $\isplit\ \rr_d\ \rr_s\ \rr_p$ &\multicolumn{3}{l}{$\idrop\ \rr$} \\
         \textsf{split} &\multicolumn{3}{l}{\textsf{discard} if $\rr$ is linear, otherwise none} \\ \hline
        \multicolumn{4}{p{.96\linewidth}}{$\ild\ \rr_d\ \rr_s$} \\
        \multicolumn{4}{p{.96\linewidth}}{\textsf{load\_linear} if $\rr_s$ is linear,
        otherwise \textsf{load} and \textsf{claim} of any linear
        capability newly reachable through $\rr_d$ after execution.
        In addition, \textsf{store\_linear} or \textsf{store} as
        appropriate if a memory location was cleared after a
        linear capability was moved out of it.} \\
        \hline
        \multicolumn{4}{p{.96\linewidth}}{$\isd\ \rr_d\ \rr_s$} \\
        \multicolumn{4}{p{.96\linewidth}}{\textsf{store\_linear} if
        $\rr_s$ is linear, otherwise \textsf{store} and
        \textsf{discard} of any linear capability moved from
        exclusive realm into shared memory.} \\ \hline
        \multicolumn{4}{p{.96\linewidth}}{$\icall\ \rr_d\ \rr_s$ and 
        $\iexcept\ \rr$} \\
        \multicolumn{4}{p{.96\linewidth}}{\textsf{send} of any linear
        capabilities transitively reachable through $\rr_s$,
        \textsf{load\_linear} as appropriate to capture LoadContext, \textsf{store\_linear} as appropriate to capture SaveContext.} \\ \hline
        \multicolumn{4}{p{.96\linewidth}}{$\ireturn\ \rr_d\ \rr_s$} \\
        \multicolumn{4}{p{.96\linewidth}}{\textsf{send} of any linear
        capabilities transitively reachable through $\rr_s$,
        \textsf{load\_linear} as appropriate to capture LoadContext, \textsf{store\_linear} as appropriate to capture ClearSealed.} \\ \hline
        \multicolumn{4}{p{.96\linewidth}}{$\iretseal\ \rr_d\ \rr_s$} \\
        \multicolumn{4}{p{.96\linewidth}}{\textsf{load\_linear} as
        appropriate to capture LoadContext, \textsf{store\_linear} as appropriate to capture ClearSealed and SaveContext (since $\iretseal$ returns a sealed capability, no transfer of linear capabilities takes place in the abstract model).} \\ \hline
    \end{tabular}
\end{table}

\subsection{Required Concrete Well-Formedness Invariants}
\noindent $\text{WF}(\mstate)$

\noindent if:\\
\indent  $\textit{caps} = \bigcup_{d \in \mathbb{N}}~\realm(\mstate, d)$\\
\indent $\forall (loc, c) \in \textit{caps}.~\text{LinearCap}(c)
\longrightarrow$\\
\indent\indent$\forall (loc', c')  \in (\textit{caps} - (loc,
c)).$\\
\indent\indent\indent$\bigcup\text{ranges}(\{(loc, c)\}) \cap \bigcup\text{ranges}(\{(loc', c')\}) = \emptyset \vee (c'.t = \tRev~\wedge)$ \\
\indent $\forall (loc, c) \in \textit{caps}.~c.t = \tRev
\longrightarrow$\\
\indent\indent$\exists (loc', c')  \in (\textit{caps} - (loc,
c)).~\bigcup\text{ranges}(\{(loc, c)\}) \cap \bigcup\text{ranges}(\{(loc', c')\}) \neq \emptyset \wedge c'.t \in \{\tSealed, \tSealedRet, \tLin\} \longrightarrow $\\
\indent\indent\indent$(\fullN \times \{(n, \tRLin)\} \cap rt) \neq \emptyset$\\

Intuitively: (1) a capability may never overlap with any other capability, unless it is non-linear and (2) every linear capability in the state must have a corresponding entry in the revocation tree.

\subsection{Statement of Correctness of Refinement Mapping}
Define a function from concrete state and domain information to abstract action.\\

$\text{Step}_{(\textit{d,$D_\textsf{sub}$})}(\mstate,\sigma) :: (\textit{act}_\textit{abs}, \textit{dom})$\\

\noindent Show that if\\
$\text{WF}(\mstate)~\wedge$\\
$\text{Refines}_\textit{(d,$D_\textsf{sub}$)}(\mstate, \textit{pstate})~\wedge$\\
$\text{Step}_{(\textit{d,$D_\textsf{sub}$})}(\mstate, \sigma) = (\textit{act}_\textit{abs},\textit{dom})~\wedge$\\
$F(\mstate, \sigma) = (\mstate^\prime, \sigma^\prime)~\wedge$\\
$\text{Refines}_\textit{(d,$D_\textsf{sub}$)}(\mstate', \textit{pstate}')$\\
then\\
$\text{WF}(\mstate')~\wedge$\\
$\textit{pstate} \hookrightarrow^{\textit{act}^\ast_\textit{abs}}_\textit{dom} \textit{pstate}'$\\

\subsection{Proof of Correctness of Refinement Mapping}
We must define our step function and prove the above.
We can proceed by case analysis on the instruction cases of our $F(\mstate, \sigma) = (\mstate^\prime, \sigma^\prime)$ and the underlying definition of Execute, defining the cases of Step inline by picking appropriate $\textit{act}_{\textit{abs}}$.
Note that in each case the currently executing domain $dom$ is given by the following, where $(s, k, \sigma^\prime) = R(\sigma)$:

$dom = \begin{cases}
    \textsf{user} & \mstate.k.d = d\\
    \textsf{sub} & \mstate.k.d \in D_\textsf{sub}\\
    \textsf{sup} & \text{otherwise}\\
\end{cases}$

\subsubsection{Trivial cases}
\hphantom{0}\\
$\imov\ \rr_d\ \rr_s$\\
$\ili\ \rr\ n$\\
$\iadd\ \rr_d\ \rr_s$ \\
$\ijnz\ \rr_d\ \rr_s$ \\
$\ilt\ \rr_d\ \rr_a\ \rr_b$\\
$\ilcc\ \rr_d\ \rr_s$\\
$\iscc\ \rr_d\ \rr_s$\\
$\iinvalid$\\

These cases do not involve any changes to (the domains of) the capabilities in the system, nor to the state of the heap memory.
Therefore $\text{WF}(\mstate')$ and $\textit{pstate} = \textit{pstate}'$ and \mbox{$\textit{act}^\ast_\textit{abs} = \epsilon$} and \mbox{$\textit{pstate} \hookrightarrow^{\epsilon}_\textit{dom} \textit{pstate}'$}

\subsubsection{Interesting cases}
\hphantom{0}\\
\noindent$\ild\ \rr_d\ \rr_s$\\

\noindent Assume \\
$\mstate = \mstatefd$\\
$\textit{pstate} = (\textit{mem}_\textit{abs}, \textit{tstate}_\textsf{user}, \textit{tstate}_\textsf{sup},  \textit{tstate}_\textsf{sub})$\\
$\mstate' = \fUpdatePC(\mstate[\Theta.k.\theta.\rr_d \mapsto
        w][mem.addr \mapsto \fMoved(w)])$ \\
$\Theta.k.\theta.\rr_s = c$\\
$\fValidCap(rt, c)$\\
$\neg\fRevoked(rt, c)$\\
$\fInBoundCap(c)$\\
$\fAccessibleCap(c)$\\
$\fReadableCap(c)$\\
$c.a = addr$\\
$mem.addr = w$ \\
(i.e. a successful concrete reduction step in the operational semantics)\\

\noindent Have $\text{WF}(\mstate')$ from assumptions+definitions (in particular, $\fMoved$ case for linear caps ensures no overlap).\\

\noindent For the second top-level proof obligation we must consider two cases:

\noindent (1) consider the case $\neg\text{LinearCap}(c)$.\\

have $\forall d.~w \notin \xrealm(\mstate,d)$ from $\text{WF}(\mstate)$ and $mem.addr = w$ and definitions.
(i.e. $w$ can't be in an exclusive realm because it falls within the range of a non-linear cap, and no linear cap can overlap with the non-linear cap)\\

moreover, have $\forall d.~\textit{addr} \notin \fworanges(\realm_w (\mstate, d))$ from $\text{WF}(\mstate)$, $\fInBoundCap(c)$, and definitions.

therefore $w = \textit{mem}_\textit{abs}[addr]$ and $\nexists\textit{dom'}.~ \text{Owns}(\textit{tstate}_{\textit{dom}'},addr)$\\

Two sub-cases:\\
\indent (1.a) $\text{LinearCap}(w)$ \\
\indent (i.e. we are loading a linear cap from non-linear memory - in this case the loaded cap and all newly transitively-reachable linear caps enter our exclusive realm)\\

In this case we must also collect the linear capabilities transitively reachable from $w$, given by $\hat{w} = \text{cl}_{\texttt{Lin}}(\mstate, \{w\})$.\\

We choose $\textit{act}^\ast_\textit{abs} = (\textsf{load}~addr~w)~(\textsf{claim}~(\text{ranges}(\hat{w})))~(\textsf{store}~addr~\text{Moved}(w))$, and must therefore additionally show that 
$\text{ranges}(\hat{w}) \notin \textit{tstate}_{\textsf{user} \cup \textsf{sup} \cup \textsf{sub}}$, from above and the definitions of Refines and $\xrealm$.\\

(1.b) otherwise we have $\neg\text{LinearCap}(w)$ and $\textit{act}^\ast_\textit{abs} = (\textsf{load}~addr~w)~(\textsf{store}~addr~\text{Moved}(w))$\\

\noindent (2) second top-level case, if $\text{LinearCap}(c)$ then $\textit{act}_\textit{abs} = (\textsf{load\_linear}~addr~w)~(\textsf{store\_linear}~addr~\text{Moved}(w))$\\

have $w \in \xrealm(\mstate,\textit{dom})$ from $\text{WF}(\mstate)$ and $mem.addr = w$ and definitions.
(i.e. $w$ must be in the current domain's exclusive realm, as it's accessed through a linear cap held in a local register)\\

moreover, have $\forall d.~\textit{addr} \notin \fworanges(\realm_w (\mstate, d))$ from $\text{WF}(\mstate)$, $\fInBoundCap(c)$, and definitions.\\

therefore $w = \textit{mem}_\textit{abs}[addr]$ and $\text{Owns}(\textit{tstate}_{\textit{dom}},addr)$\\

therefore $\textit{pstate} \hookrightarrow^{\textit{act}_\textit{abs}}_\textit{dom} \textit{pstate}'$\\

QED\\

\noindent$\isd\ \rr_d\ \rr_s$\\
\noindent Assume:\\
$\mstate = \mstatefd$\\
$\textit{pstate} = (\textit{mem}_\textit{abs}, \textit{tstate}_\textsf{user}, \textit{tstate}_\textsf{sup},  \textit{tstate}_\textsf{sub})$\\
$\mstate' = \fUpdatePC(\mstate[mem.addr \mapsto
    w][\Theta.k.\theta.\rr_s \mapsto \fMoved(w)][\Theta.k.\theta.\rr_d \mapsto
    \fUpdateCursor(c)])$ \\
$\Theta.k.\theta.\rr_d = c$\\
$\fValidCap(rt, c)$\\
$\fInBoundCap(c)$\\
$\fAccessibleCap(c)$\\
$\fWritableCap(c)$\\
$c.a = addr \land \Theta.k.\theta.\rr_s = w$ \\

\noindent have $\text{WF}(\mstate')$ from definitions (in particular, $\fMoved$ case for linear caps ensures no overlap).\\

\noindent For the second top-level proof obligation we must consider two cases:

\noindent (1) first, consider the case $\neg\text{LinearCap}(c)$\\

have $\forall d.~w \notin \xrealm(\mstate,d)$ from $\text{WF}(\mstate)$ and $mem.addr = w$ and definitions.
(i.e. $w$ can't be in an exclusive realm because it falls within the range of a non-linear cap, and no linear cap can overlap with the non-linear cap)\\

moreover, have $\forall d.~\textit{addr} \notin \fworanges(\realm_w (\mstate, d))$ from $\text{WF}(\mstate)$, $\fInBoundCap(c)$, and definitions.

therefore $w = \textit{mem}_\textit{abs}[addr]$ and $\nexists\textit{dom'}.~ \text{Owns}(\textit{tstate}_{\textit{dom}'},addr)$\\

Two sub-cases:\\
\indent (1.a) $\text{LinearCap}(w)$ \\
\indent (i.e. we are storing a linear cap into non-linear memory - in this case the stored cap and all transitively-reachable linear caps leave our exclusive realm)\\

In this case we must also collect the linear capabilities transitively reachable from $w$, given by $\hat{w} = \text{cl}_{\texttt{Lin}}(\mstate, \{w\})$.\\

We choose $\textit{act}^\ast_\textit{abs} = (\textsf{store}~addr~w)~(\textsf{discard}~(\text{ranges}(\hat{w})))$.\\

\indent (1.b) otherwise we have $\neg\text{LinearCap}(w)$ and $\textit{act}^\ast_\textit{abs} = \textsf{store}~addr~w$\\

\noindent (2) second top-level case, if $\text{LinearCap}(c)$ then $\textit{act}_\textit{abs} = \textsf{store\_linear}~addr~w$\\

have $w \in \xrealm(\mstate,\textit{dom})$ from $\text{WF}(\mstate)$ and $mem.addr = w$ and definitions.
(i.e. $w$ must be in the current domain's exclusive realm, as it's held in a local register)\\

moreover, have $\forall d.~\textit{addr} \notin \fworanges(\realm_w (\mstate, d))$ from $\text{WF}(\mstate)$, $\fInBoundCap(c)$, and definitions.\\

therefore $w = \textit{mem}_\textit{abs}[addr]$ and $\text{Owns}(\textit{tstate}_{\textit{dom}},addr)$\\

therefore $\textit{pstate} \hookrightarrow^{\textit{act}_\textit{abs}}_\textit{dom} \textit{pstate}'$\\

QED\\

\noindent$\iseal\ \rr$\\

This case is trivial, with $\text{WF}(\mstate')$ and $\textit{pstate} = \textit{pstate}'$ and \mbox{$\textit{act}^\ast_\textit{abs} = \epsilon$}, however it's worth noting \textit{why} explicitly - sealing a capability does not alter the domains of any capabilities reachable through it.
Note that a domain sealed inside a linear capability has no control in general over when the capability is unsealed, so the abstract model simply over-approximates that all sealed capabilities are immediately available to the sealed domain.\\

\noindent$\icall\ \rr_d\ \rr_s$\\

\noindent Assume:\\

\noindent $\mstate = \mstatefd$\\
$\textit{pstate} = (\textit{mem}_\textit{abs}, \textit{tstate}_\textsf{user}, \textit{tstate}_\textsf{sup},  \textit{tstate}_\textsf{sub})$\\
$\mstate' = \mstate[\Theta.k \mapsto (regs^\prime, d)][mem \mapsto
mem^\prime]$ \\
$\Theta.k.\theta.\rr_d = c_i$\\
$\Theta.k.\theta.\rr_s = w$\\ 
$\fValidCap(rt, c_i)$\\
$c_i = (\tSealed(d), b_i, e_i, a_i, p_i, n_i)$\\
$b_i + M + 4 \leq e_i$\\
$regs^\prime = \fLoadContext(\Theta.k.\theta, b_i, mem)[\rret \mapsto c_i[t \mapsto
    \tSealedRet(\mstate.k.d, \rr_d)]][\rr_1 \mapsto w]$\\
$mem^\prime = \fSaveContext(mem, b_i, \Theta.k.\theta[\rr_s \mapsto
    \fMoved(w)][\rr_d \mapsto 0])$\\

We must show that the capability ownership changes in the abstract
domains are congruent with the changes in the concrete domains.
First, note that the only way that a capability switches domains as a
result of executing $\icall$ is if the argument register $\rr_s$
contains a capability, since the cap in $\rsc$ is linear, so
$\fSaveContext$ will only move capabilities within the same domain.
Moreover, $\fLoadContext$ loads capabilities into the registers from
$c_i$, a sealed capability with domain $d$, but the currently
executing domain is also switched to $d$, so no transfer takes place.
If $w$ is a linear capability, we define $\hat{w} =
\text{cl}_{\texttt{Lin}}(\mstate, \{w\})$, otherwise we define
$\hat{w} = \epsilon$.
Then $\textit{act}^\ast_\textit{abs} = \textsf{send}~(\text{ranges}(\hat{w}))~d$, plus
appropriate \textsf{load/store} actions to handle $\fSaveContext$ and
$\fLoadContext$ (trivial details of these omitted).

To show $\textit{pstate} \hookrightarrow^{\textit{act}_\textit{abs}}_\textit{dom} \textit{pstate}'$, we must show that the capability ownership changes in the abstract domains are congruent with the changes in the concrete domains.
We can show this by observing that all registers in the current domain are saved to a sealed capability for the same domain (i.e. no domain transfer), except $\rr_d$ and $\rr_s$ which are cleared appropriately, with the contents of $\rr_d$ being loaded into the newly-executing domain, and the transfer of the contents of $\rr_s$ via $\rr_1$ captured by our choice of \textsf{send} action.\\

\noindent$\ireturn\ \rr_d\ \rr_s$\\

Essentially the same as $\icall$, with appropriate memory actions for the LoadContext, ClearSealed, and SaveContext operations in the concrete semantics.\\

\noindent$\iretseal\ \rr_d\ \rr_s$\\

Essentially the same as $\icall$, with appropriate memory actions for the LoadContext, ClearSealed, and SaveContext operations in the concrete semantics.\\

\noindent$\idrop\ \rr$\\

Maps straightforwardly to the \textsf{discard} operation.\\


\noindent$\irevoke\ \rr$\\

\noindent Assume:\\
$\mstate = \mstatefd$\\
$\textit{pstate} = (\textit{mem}_\textit{abs}, \textit{tstate}_\textsf{user}, \textit{tstate}_\textsf{sup},  \textit{tstate}_\textsf{sub})$\\
$\mstate' = \fUpdatePC(\mstate[\Theta.k.\theta.\rr \mapsto
    c^\prime][rt \mapsto rt^\prime])$\\
$\Theta.k.\theta.\rr = c \land \fValidCap(rt, c)$\\
$c = (\tRev, b, e, a, p, n)$\\
$rt^\prime = \fReparent(rt, n, \nnull)$\\
$c^\prime = \begin{cases}
    (\tLin,
        b, e, a, p, n) & (\fullN \times \{(n, \tRLin)\} \cap rt) =
        \emptyset \\
    (\tUninit, b, e, b, p, n) & \text{otherwise}
    \end{cases}$\\

Have $\text{WF}(\mstate')$ from assumptions+definitions (in particular, following the definition of Reparent to ensure the revocation tree is updated correctly).\\

Two cases:

First, $(\fullN \times \{(n, \tRLin)\} \cap rt) =
        \emptyset$\\
        
We pick $\textit{act}^\ast_\textit{abs} = \textsf{claim}~(\text{ranges($\{c\}$)})$\\

Must show that $\text{ranges($\{c\}$} \notin \textit{tstate}_{\textsf{user}\cup\textsf{sub}\cup\textsf{sup}}$. This follows from $\text{WF}(\mstate)$ and the above (i.e. since there is no entry in the revocation tree, by the definition of \text{WF} there can be no linear capability in any exclusive realm that overlaps with $c$).\\
        
Second, $(\fullN \times \{(n, \tRLin)\} \cap rt) \neq
\emptyset$\\

We pick $\textit{act}^\ast_\textit{abs} = \textsf{revoke}~(\text{ranges($\{c\}$)})$\\

To establish $\textit{pstate} \hookrightarrow^{\textit{act}_\textit{abs}}_\textit{dom} \textit{pstate}'$, we must show\\

(a) $(\textit{dom} = \textsf{sub} \longrightarrow \textit{cap} \notin \textit{tstate}_\textsf{user})$\\
\indent (b) $(\textit{mem}_\textit{abs}' = \textit{mem}_\textit{abs}[\textit{cap} := \textsf{uninit}])$\\

(a) follows from the invariant $\text{Sub}(\mstate,D_{\textsf{sub}})$ that is contained within Reifies, which enforces that \textit{c} cannot overlap with any user domain capability.\\

(b) holds by the definition of Reifies, as we know in this case that $c'.t = \tUninit$\\

\noindent$\idelin\ \rr$\\

Essentially the same as the $\tLin$ case of $\idrop$.\\

\noindent$\imrev\ \rr_d\ \rr_s$\\

This case is trivial, with $\text{WF}(\mstate')$ and $\textit{pstate} = \textit{pstate}'$ and \mbox{$\textit{act}^\ast_\textit{abs} = \epsilon$}, however it's worth noting \textit{why} explicitly - the abstract model permits domains to revoke at any time, essentially over-approximating that they have already minted all of the revocation capabilities that they possibly can (with reference to the restrictions on the subordinate environment).\\

\noindent$\itighten\ \rr_d\ \rr_s$\\

This case is trivial, with $\text{WF}(\mstate')$ and $\textit{pstate} = \textit{pstate}'$ and \mbox{$\textit{act}^\ast_\textit{abs} = \epsilon$}, however it's worth noting \textit{why} explicitly - the abstract model over-approximates that owning a linear capability grants full permissions to the relevant underlying memory.
This is acceptable since we don't allow both a separate read-only and write-only linear cap for the same memory location to exist at once.\\

\noindent$\isplit\ \rr_d\ \rr_s\ \rr_p$\\

Maps straightforwardly to the \textsf{split} abstract action.\\

\noindent$\ishrink\ \rr_d\ \rr_b\ \rr_e$\\

Maps straightforwardly to the \textsf{shrink} abstract action.\\

\noindent$\iinit\ \rr$\\

Another trivial case, since the abstract model over-approximates that uninitialized capabilities confer arbitrary access to any memory that is not \textsf{uninit}.\\

\noindent$\iexcept\ \rr$ \\

The concrete model defines exceptions in terms of calls, so identical to the call case above.\\

\subsection{Auxiliary Definitions}

\begin{definition}[Capability closure]
    For any $\mstate \in \sState$, $S \subseteq \text{DLoc} \times
    \sCap$ and $T \subseteq
    \sCapType$, we inductively define the capability closure of $S$ with respect
    to $T$ at $\mstate$, denoted as $\text{cl}_T(\mstate, S)$, as
    follows
    \begin{itemize}
        \item $\{ s \mid s \in S \wedge s.t \in T \} \subseteq \text{cl}_T(\mstate, S)$
        \item If $\fValidCap(\mstate.rt, \mstate.mem.u) \land \mstate.mem.u.t
        \in T \land b \leq u < e \land c^\prime = (t, b, e, a, p, n)$
        where $(loc, c^\prime) \in \text{cl}_T(\mstate, S)$, then
        $(\sDLocMem(u), \mstate.mem.u) \in \text{cl}_T(\mstate, S)$
    \end{itemize}
\end{definition}

\begin{definition}[Execution context]
For any domain $d \in \mathbb{N}, \mstate \in \sState$, its execution
    context $\ctx (\mstate, d)$ is defined as
\begin{equation}
    \ctx(\mstate, d) = \begin{cases}
        \{(\sDLocReg(k, u), \mstate.\Theta.k.\theta.u) \mid u \in \fullN\} & \exists k \in
        \fullN, \mstate.\Theta.k.d = d  \\
        \{(\sDLocMem(u, \mstate.mem.u)) \mid (r, u) \in \fSealedRegs(c) \} & \exists c \in
        \text{Dom}(\mstate.mem) \cup \bigcup_{k \in \fullN}
        \text{Dom}(\mstate.\Theta.k.\theta), \fValidCap(c) \land \\
        & c.t \in \{\tSealed(d), \tSealedRet(d)\} \\
        \emptyset & \text{otherwise}.
    \end{cases}
\end{equation}
\end{definition}

\begin{definition}[Realm]
For any domain $d \in \mathbb{N}, \mstate \in \sState$, its realm
    is defined as 
    \begin{equation}
        \realm(\mstate, d) = \text{cl}_T(\mstate, S),
    \end{equation}
    where
    \begin{equation}
        T = \{\tLin, \tNon\},
    \end{equation}
    and
    \begin{equation}
        S = \{(loc, c) \mid \fValidCap(\mstate.rt, c) \land c.t \in T,
        (loc, c) \in \ctx(\mstate, d)\}.
    \end{equation}
\end{definition}

\begin{definition}[Exclusive realm]
For any domain $d \in \mathbb{N}, \mstate \in \sState$, its exclusive realm is defined as
    \begin{equation}
        \xrealm(\mstate, d) = \text{cl}_{T_x}(\mstate, S_x),
    \end{equation}
    where
    \begin{equation}
        T_x = \{\tLin\},
    \end{equation}
    and
    \begin{equation}
        S_x = \{(loc, c) \mid \fValidCap(\mstate.rt, c) \land c.t \in
        T_x,
        (loc, c) \in \ctx(\mstate, d)\}.
    \end{equation}
\end{definition}

\deffunc{ranges}

We can easily define a function that converts realms and exclusive
realms to sets of ranges:
\begin{equation}
    \franges(\realm) = \left\{\{b, b + 1, \cdots, e - 1\} \mid (loc, (t, b,
    e, a, p, n)) \in \realm \right\}.
\end{equation}

This following is the counterpart of a realm for write-only memory
(for uninitialized capabilities).
\begin{definition}[Write-only realm]
For any $d \in \fullN, \mstate \in \sState$, the write-only realm of $d$ in
    $\mstate$ is defined as 
\begin{equation}
    \realm_w (\mstate, d) = \{(loc, c) \mid (loc, c) \in \hat{\realm}(\mstate, d),
    \fValidCap(\mstate.rt, c), c.t = \tUninit \},
\end{equation}
where
\begin{equation}
    \hat{\realm}(\mstate, d) = \{(\sDLocMem(u), \mstate.mem.u) \mid u \in
    range \in \franges(\realm(\mstate, d)) \} \cup \ctx(\mstate, d).
\end{equation}
\end{definition}

\begin{definition}[Exclusive write-only realm]
For any $d \in \fullN, \mstate \in \sState$, the exclusive write-only realm of $d$ in
    $\mstate$ is defined as 
\begin{equation}
    \xrealm_w (\mstate, d) = \{(loc, c) \mid (loc, c) \in \hat{\xrealm}(\mstate, d),
    \fValidCap(\mstate.rt, c), c.t = \tUninit \},
\end{equation}
where
\begin{equation}
    \hat{\xrealm}(\mstate, d) = \{(\sDLocMem(u), \mstate.mem.u) \mid u \in range \in
    \franges(\xrealm(\mstate, d)) \} \cup \ctx(\mstate, d).
\end{equation}
\end{definition}

We can define a function that converts a write-only or exclusive
write-only realm to ranges:
\begin{equation}
    \fworanges(\realm_w) = \left\{\{a, a + 1, \cdots, e - 1\} \mid
    (loc, (t,
    b, e, a, p, n)) \in \realm_w \right \}.
\end{equation}

\end{document}